\documentclass[12pt]{article}
\usepackage{booktabs} 
\usepackage{pdflscape} 
\usepackage{multirow} 
\usepackage{bbm} 
\usepackage{comment}
\usepackage{amsmath}
\usepackage{amssymb}
\usepackage{enumitem}
\usepackage{emptypage}
\usepackage{marginnote}
\usepackage[margin=1in]{geometry}
\usepackage{microtype}
\usepackage{url}
\usepackage{graphicx}
\usepackage{subfig}
\usepackage{enumitem}
\usepackage{floatrow} 
\usepackage{tikz}
\usepackage[section]{placeins}
\usepackage{lscape}
\usepackage{array}
\usepackage{pgfplots}
    \allowdisplaybreaks
\usepackage{natbib}
\usepackage{authblk}
\usepackage{adjustbox}
\usepackage{bookmark} 
\usepackage{cleveref}
	\crefname{appsec}{Appendix}{Appendices}
\usepackage{color,soul} 
\usepackage{floatpag} 
\usepackage[singlelinecheck=false]{caption} 
\usepackage{datetime}

\newcommand{\figspath}{./}

\newcommand{\killpunct}[1]{}

\renewcommand{\citet}[1]{\citep{#1}}

\newdateformat{mydate}{\monthname[\THEMONTH] \THEYEAR}

\title{Crowded trades, market clustering, and price instability}

\author[1]{Marc~van~Kralingen}
\author[2]{Diego~Garlaschelli}
\author[3]{Karolina~Scholtus}
\author[4]{Iman~van~Lelyveld\thanks{\scriptsize Views expressed are those of the authors and do not necessarily reflect official positions of De Nederlandsche Bank or Aegon N.V.. We would like to thank colleagues at the De Nederlandsche Bank, VU Amsterdam, Wouter van Bronswijk (AFM), Richard Verhoef (AFM), Dick van Dijk (EUR), and participants at the Risklab/BoF/ESRB -- Conference on Systemic Risk Analytics (2016), Bristol Banking and Financial Intermediation Workshop (2017), the 25th International Panel Data Conference (2019), for input received. Naturally all errors are ours. Corresponding author: iman.van.lelyveld@dnb.nl -- +31652496159.}}
\affil[1]{\small Aegon N.V}
\affil[2]{\small Lorentz Institute for Theoretical Physics, Leiden University}
\affil[3]{\small Erasmus University Rotterdam}
\affil[4]{\small De Nederlandsche Bank, VU Amsterdam}

\begin{document}
\mydate
\maketitle
\thispagestyle{empty}

\begin{abstract}
\noindent Crowded trades by similarly trading peers influence the dynamics of asset prices, possibly creating systemic risk. We propose a market clustering measure using granular trading data. For each stock the clustering measure captures the degree of trading overlap among any two investors in that stock. We investigate the effect of crowded trades on stock price stability and show that market clustering has a causal effect on the properties of the tails of the stock return distribution, particularly the positive tail, even after controlling for commonly considered risk drivers. Reduced investor pool diversity could thus negatively affect stock price stability. \vspace{12pt}
\end{abstract}

\newpage
\setcounter{page}{1}
\section{Introduction}
	\label{section:introduction}
This paper studies the effect of market clustering on price instability. We define market clustering as the degree to which groups of investors trade similarly. For each stock our market clustering model measures the degree of trading overlap among any two investors that trade that particular stock. In general, stock prices are thought to adjust continuously to changes in the fundamental value of the stocks. The reactions of investors to new information determine the adjustments of prices and the resulting price dynamics. Market clustering, however, cannot be observed by individual investors and its effect on price dynamics can thus unfold unexpectedly.

Market clustering can be seen as a measure of the homogeneity of the investors' pool. Reduced diversity of the investors' pool, i.e., when the investors show similar trading behaviour, means that coincidental overlap of trading strategies is more likely and overlap of trades increase the chance of crowded trades and overreactions, reflected in price fluctuations. The use of large-scale granular trading data and a novel complex network method enables us to study the effect of market clustering on price fluctuations directly. To the best of our knowledge, this is the first direct empirical investigation of the relation between market clustering and price fluctuations on individual stock level.

Studying the empirical relation between market clustering and price instability is relevant from both an academic and a supervisory point of view. Firstly, the existing empirical literature on the topic focuses on only indirect measures of group behavior: overlapping portfolio's \citet{anton2014, bruno_portfolio_2018}, similarities in performance dynamics \citet{pojarliev2010, kinlaw_crowded_2018}, dynamics of the number of owners per stock \citet{hong2013}, 
or buyer and seller volume imbalance \citet{yang_individual_2016, jia_disagreement_2017}. The suggestion that price fluctuations originate from uncoordinated or inefficient interaction among investors, seems obvious, but due to limited data and lack of suitable methods such effects have not yet been investigated directly. 

Secondly, knowledge about the implications of market clustering is relevant for regulators, since market clustering can be an amplifying spillover channel for asset price fluctuations. The general implication of a causal relation between market clustering and price instability is that trading patterns through which investors react to incentives, matter for the efficiency of price discovery. Although this research focuses on the effect of market clustering on single stocks, market clustering might be a channel of volatility spillovers, because portfolio adjustments concerning other stocks in reaction to an initial price shock are more likely to overlap as well in a clustered trading environment. Therefore, market clustering might not only be a source of price instability, but also a channel of volatility spillovers, eventually resulting in correlated price jumps. In that case, market clustering would foster systemic risks. Market clustering might be an example of an existing market structure that can amplify seemingly unimportant events into wide-spread market volatility. In case market clusters coincide with otherwise interconnected institutions, for example banks, common asset devaluation can be a crucial default contagion channel, as suggested in recent interdisciplinary research \citet{tsatskis2012,Glasserman2016JEL,levy2015}.

Market clustering is expected to cause price shocks, because it amplifies the effect of existing sources of price fluctuations. More specifically, market clustering is expected to increase the chance of price shocks in two different situations: Firstly, when the order deluge due to the group behaviour overwhelms the supply \citet{stein2009,BraunMunzinger2018} and, secondly, when the supply is thin due to the homogeneity of the investors' pool, i.e. a lack of liquidity at one side of the order book \citet{weber2006}. In both situations market clustering increases the chance that the demand exceeds the supply, either in buy or sell orders.

We start our investigation of the influence of trading patterns by studying the relation between market clustering and the price dynamics of individual stocks. Our market clustering measure is unique in the sense that it quantifies two aspects of group behavior: clustering and crowdedness.
We define price instability as an increase of the number of sharp price fluctuations, such that the tails of the log return distribution are heavier. Specifically, we investigate whether there is a causal relation between market clustering and the skewness, kurtosis, tail indices, positive and negative outlier counts, changes in downside risk and upside gains.

The analysis of trading patterns depends on the ability to distinguish to what extent the patterns are the result of higher order features like group behaviour instead of lower order properties. In this research, we represent stock trading by a complex bipartite network (a two-layer network with non-trivial topological features). The two layers are the investors and the stocks. The links between the layers are the trades during a particular time period. We use the maximum entropy principle to reconstruct the network using only lower order properties, in this case the degree sequences (i.e., the number of investors per stock and the number of stocks per investors). Most apparent network patterns can be explained by the lower order properties. Observations that deviate from the reconstructed network are indications of higher order trading patterns. This method builds on recent research on the topological structures of the world trade network, showing that the lowest order property that explains the occurrence of tryadic structures (three mutually trading countries) is the reciprocity of the bilateral (dyadic) trading relations \citet{squartini2012}.

Our source data consist of granular trade-by-trade records of Dutch banks and investment funds. These data are reported under the Markets in Financial Instruments Directive (MiFID). The data available to us contain all the transactions in stocks and bonds traded by all Dutch banks and investments firms (approximately 50). These trades are either conducted as an agent or for own account. The set of investors per stock is incomplete, since trades are only reported in this data set if a Dutch bank or investment firm is involved and hence we do not observe trades between two foreign parties.

The results indicate that the prices are less (more) stable for high (low) market clustering. We find evidence for a consistent and robust positive relation between market clustering and the kurtosis of the log return distribution. Clustering thus seems to related to large price movements. Furthermore, we find a relation between market clustering and the tail index and outlier count for the positive tail, but interestingly not for the negative tail.

We use the data limitation that we do not observe trading among foreign investors to mimic an experimental research design and study the causality of the relation between market clustering and price instability. Per stock we measure what percentage of its turnover is traded by investors included in the MiFID data set and compare the results for stocks that are mainly traded by included investors ("treatment group") with the results for stocks that are mainly traded by investors elsewhere ("control group"). Here we find evidence for causality as our results do not hold for stocks mostly traded by non-Dutch investors.

Finally, we examine market clustering and price instability in a dynamic panel data framework. These models show that clustering is a persistent process, affected by market conditions but not by stock return momentum or fundamental variables. The only stock-related variables that matter are liquidity and market capitalization. Higher illiquidity in low volatility periods leads to higher clustering scores while the coefficient for illiquidity in a high volatility state is insignificant. The model confirms the possibility that crowded trades lead to fire sales as less liquid stocks are traded more in downward markets. The relation between market capitalization and market clustering supports the presence of flight-to-safety within equities in turmoils and investor efforts to reap a size premium in upward markets. Thus the results are consistent with multiple equity market phenomena.

When we investigate the drivers of changes in Value-at-Risk (VaR) and Value-at-Luck (VLuck; upside potential, measured as VaR but for the right side of the return distribution), we find that our proposed clustering measure has explanatory power beyond other well-known variables. Our conditioning variables include practically all the variables suggested in the literature (i.e., market factor, book-to-market, dividend yield, size, Amihud (2002) liquidity measure, momentum, market conditions). The findings confirm that stock's involvement into crowded trades lead to larger price fluctuations. The effect is stronger for the positive tail (VLuck) and consistent with results from group comparisons. For the negative tail market clustering causes price instability during financial turmoil, but not during calm periods. The dynamic panel data framework we employ conveniently accounts for endogeneity, making causality claims reliable.

The setup of the remainder of the paper is as follows. First, we provide a brief overview of the relevant literature. Then we turn to a description of the data, followed by an explanation of the method to measure market clustering we developed. To the best of our knowledge, both the data and the method are new contributions to the literature. We then describe our results and close with a discussion.

\section{Literature review}
	\label{section:theory}

The literature studying price dynamics is rich and can be classified in many ways. Our focus here is on joint trading affecting the market in such a way that it is no longer capable to perform two key functions: efficient price discovery and providing liquidity \citet{OHara2003}. Several related strands of the literature shed light on this important issue covering 1) similar shocks on the funding side, 2) overlapping portfolios, 3) exogenous requirements, 4) market microstructure design issues, and 5) complexity models.

First, some argue that participants in the market face very similar funding shocks or, more generally, that investment needs or beliefs are highly correlated. This affects prices because leverage cycles result in fat tails \citet{Thurner2012}. For instance, \cite{gorban_trading_2018} suggest a continuous-time model where beliefs of strategic informed traders about crowdedness of trades and strategies in the market can lead to reduced liquidity on supply side and lower market depth.

Second, given the investment needs and outlook, investors will have accumulated a portfolio of assets that might to some degree be overlapping. With homogenous agents and perfect information, all portfolios will approach the market portfolio.\footnote{See, however, \cite{Wagner2011}, who develops a model where investors purposefully choose divergent portfolio's to avoid being forced to join a market wide fire-sale.} In practice, investors are heterogeneous and information is uncertain and not freely available, thus investors will have portfolios that overlap only partly.
This does not limit itself to liquid investments but also applies to longer term and less liquid exposures such as in the syndicated loan market \citet{Cai2018}.

Common asset holdings have attracted considerable attention, especially in the context of fire-sale spill-overs and cascade dynamics \citet{Caccioli2014,Caccioli2015,Greenwood2015}. Not surprisingly, studies find that more commonality in investments increase systemic risk with an exception to \cite{barroso_institutional_2018}, who discover no evidence of the relation between momentum crashes and institutional crowding. \cite{gualdi_statistically_2016} show that portfolio overlapping on aggregate level increased slowly before the 2008 crisis, reached a peak at the start of the crisis and then triggered fire sales. Moreover, network effects are generally important (although \cite{Glasserman2015} come to the opposite conclusion). Theoretical work has evolved from analysing the effect of fire sales on a single portfolio and a single asset \citet{Brunnermeier2005} to continuous time models with endogenous risk and spill-over from fire sales across multiple assets and multiple portfolios \citet{Cont2014}. Empirical (stress test) exercises assess how relevant such contagion effects are in practice. The results are highly dependent on the financial system considered \citep[for example,][]{Cont2016}.

A third area of the literature relevant for our analysis highlights fire sales caused by an exogenous requirement. Note that fire sales are forced sales in stressed markets under unfavourable terms and are very different compared to regular buying and selling to adjust a portfolio. External requirements are often set by regulators to safeguard sufficient buffers for various risks (credit risk -- using both risk weighted and risk insensitive measures (i.e. leverage ratios), counterparty credit risk, or liquidity risk \citet{Bluhm2014,Cont2016,Ellul2011,Thurner2012,Aymanns2015a}.

Regulatory requirements often imply cliff effects since breaching certain thresholds come with costs. External demands leading to forced sales can sometimes also come from other market participants. For example, counterparties can call for margin. In particular, central clearing parties can require substantial margins to be delivered at very short notice (\cite{Glasserman2018ManScience}).

Fourthly, there is an established literature on mispricing because of market microstructure design and crowded trades \citet{madhavan2000,stein2009,BraunMunzinger2018}. Sometimes investors are prone to herding \citet{Pedersen2009}, at other times, speculators try to manipulate prices by rapidly submitting orders to drive up prices.

Finally, we develop and apply complexity models -- as recently advocated by \cite{Battiston2016}. Network theory in general has many applications in finance \citet{Glasserman2016JEL} and complex network theory offers reconstruction procedures and null models based on a maximisation of entropy \citet{newman2004}. Such models have been applied to the world trade network \citet{squartini2012} and banking networks \citet{Squartini2011e, Squartini2013, huang_cascading_2013}.\footnote{A slightly different type of network emerges from order optimisation as studied by \cite{Cohen-Cole2014}. In studying the DOW and the S\&P e-mini futures, they show that in these entirely electronic markets economically meaningful networks emerge. This happens despite the fact that the interjection of an order-matching computer makes social interaction impossible.}. In the method we develop here -- to be elaborated on below -- we incorporate the distribution of the number of links (degree distribution) but otherwise our expectation (or null model) is as random as possible.

To clarify our approach to crowded trading, we present a graphical representation of market clustering in \Cref{fig:m_theory}. The homogeneity of the trading behaviour of the investors' pool per stock is then reflected in the market clustering measure that we will define below in \Cref{eq:m}. In the most extreme case, the market breaks up into distinct submarkets, consisting of groups of investors that trade only in particular stocks which are only traded by those groups. Incorporating the effect of clustering into the measure on individual stock level is what sets our research apart from other crowdedness measures intended for individual stocks. For example, \cite{yang_individual_2016} differentiate between seller and buyer initiated crowded trades per stock. Their measure is based on trading volume data and thus does not reflect the (unobserved) interactions among investors. The same applies to the quarterly measure derived from mutual funds holding data in \cite{zhong_impact_2017}. The stocks that are largely held by actively managed mutual funds are classified as overcrowded but the tendency of particular stock's owners to trade with each other is not taken into account.

\begin{figure}[!htb]
\centering
\begin{adjustbox}{max width=\textwidth}
\begin{tikzpicture}



    \node[draw=none](f) at (5,7.5) {Random trades};

    \foreach \x [count=\xi] in {1.5,2.1,...,11.2}
    	\node[draw,circle,fill=black,inner sep=.7mm](s\xi) at (\x,4.5){};
    \foreach \x [count=\xi] in {1.5,3.1,...,11.2}
    	\node[draw,circle](f\xi) at (\x,7){};
    \node[draw=none](f) at (0,7) {Firms};
    \node[draw=none](s) at (0,4.5) {Stocks};

    \draw[-,line width=0.5mm] (f) to [out=-90,in=90]     node[draw,rectangle,fill=white,anchor=center]{Trades} (s);


    \draw[->,line width=.5mm,dotted] (s) to [in=-140,out=140] node[draw,rectangle,fill=white,anchor=east]{Incentives}(f);

    \foreach \x [count=\xi] in {3,8,14,17}
    	\draw[-] (f1) to (s\x);
    \foreach \x [count=\xi] in {1,4,7,15,16}
    	\draw[-] (f2) to (s\x);
    \foreach \x [count=\xi] in {2,5,13}
    	\draw[-] (f3) to (s\x);
    \foreach \x [count=\xi] in {2,5,6,7,8,9,10}
    	\draw[-] (f4) to (s\x);
    \foreach \x [count=\xi] in {2,3,5,11,12,14,17}
    	\draw[-] (f5) to (s\x);
    \foreach \x [count=\xi] in {1,3,11,13}
    	\draw[-] (f6) to (s\x);
    \foreach \x [count=\xi] in {6,9,16}
    	\draw[-] (f7) to (s\x);


    \node[draw=none](f) at (5,4) {Clustered trades};
    	
    \foreach \x [count=\xi] in {1.5,2.1,...,11.2}
    	\node[draw,circle,fill=black,inner sep=.7mm](s\xi) at (\x,1){};


    \foreach \x [count=\xi] in {1.5,3.1,4.7}
      	\node[draw,circle,fill=blue](f\xi) at (\x,3.5){};
   	\node[draw,circle,fill=white](f4) at (6.3,3.5){};
    \node[draw,circle,fill=white](f5) at (7.9,3.5){};
    \node[draw,circle,fill=red](f6) at (9.5,3.5){};
    \node[draw,circle,fill=red](f7) at (11.2,3.5){};


    \node[draw=none](f) at (0,3.5) {Firms};
    \node[draw=none](s) at (0,1) {Stocks};

    \draw[-,line width=0.5mm] (f) to [out=-90,in=90] node[draw,rectangle,fill=white,anchor=center]{Trades} (s);
    \draw[->,line width=.5mm,dotted] (s) to [in=-140,out=140] node[draw,rectangle,fill=white,anchor=east]{Incentives}(f);

    \foreach \x [count=\xi] in {1,2,3,5}
    	\draw[-] (f1) to (s\x);
    \foreach \x [count=\xi] in {1,2,3,4,5}
    	\draw[-] (f2) to (s\x);
    \foreach \x [count=\xi] in {2,3,5}
    	\draw[-] (f3) to (s\x);
    \foreach \x [count=\xi] in {6,7,8,9,11,12,13}
    	\draw[-] (f4) to (s\x);
    \foreach \x [count=\xi] in {6,7,8,9,10,11,13}
    	\draw[-] (f5) to (s\x);
    \foreach \x [count=\xi] in {14,15,16,17}
    	\draw[-] (f6) to (s\x);
    \foreach \x [count=\xi] in {14,16,17}
    	\draw[-] (f7) to (s\x);
\end{tikzpicture} 
    \end{adjustbox}
    \caption{Market clustering in a bipartite network representation. \\
    {\footnotesize The nodes in the top layer represent the firms and the bottom layer represents the stocks. The links between the layers represent all trades during a certain time period. Each line is a trade of the connected firm in the connected stock. In the top network, all trades are randomly distributed over the firms and stocks. The bottom trade network shows market clustering: Groups of firms trade in separated groups of stocks, while these stocks are traded only by these particular firms, which results in three distinct market clusters. The number of trades per firm and per security is the same for both random trades and the clustered trades example.}}
    \label{fig:m_theory}
\end{figure}

In general, peers trading similarly are likely to share common features, i.e., in case the group of investors that trade in a stock is very similar, then trading behaviour might be similar, too. In our current analysis we abstract from what drives common trading. We are thus agnostic as to whether the order flows are driven by, for example, adjustments due to common asset holdings, (too) similar investment views, or shared regulatory constraints. Note that we do investigate what makes a particular stock attractive for involvement into clustered trades and how that depends on market conditions.

\section{Methodology}
	\label{section:method}

In this section we will first discuss our novel contribution: how to define a metric for homogeneous trading. Then follows the definition of price instability and the cross-sectional comparison framework to assess the relation between clustering and price instability. Finally, we present a dynamic panel data model. We implement the latter in order to investigate the drivers of our newly defined measure as well as to show that it has additional explanatory power over and above well established covariates in models for downside risk and upside potential.

\subsection{Measuring homogeneous trading}\label{subsection:homogeneousTrading}

Our first goal here is to define a measure of similar or homogeneous trading behaviour. This indicator will then be linked to the measures of price instability to investigate whether higher order patterns affect price formation. The nexus of trades between firms and stocks is complex and exhibits both lower and higher order network properties. Lower order properties, such as the liquidity of a particular stock, have been researched extensively and are key determinants of price dynamics. Lower order properties can be seen as the exogenous causes of price instability and their effects on price dynamics are direct and undelayed.

However, we focus on whether the market microstructure conceals particular grouping of trades that disturb the efficiency of the market. Particular ordering of the trades, resulting in higher order patterns, can function as endogenous cause of price instability. Crucially, these market features are unobservable to the investors and their effects on prices can unfold unexpectedly. Such effects have not been investigated by use of granular trading data, because suitable methods had yet to developed and the data have been largely unavailable.

We develop a method that incorporates the information encoded in the number of unique investors per security and the number of unique, traded securities per firm. Our method results in a probability of a link for each security firm pair and, consequently, for combinations of links (i.e. 'motifs'). This enables us to compare the observed trading network to the network based on only the degree sequences. The deviations from the network based on only the degree sequences are indications for higher order patterns such as peers clustering in the same (type of) stock.

To identify market clustering, we need the observed values and the expected values based on the benchmark model. The quantity that represents the market clustering of security $s$ during month $t$ is
\begin{equation}
m_{s,t} = \frac{M_{s,t}}{\langle M_{s,t}\rangle^*}-1,\label{eq:m}
\end{equation}
where $M_{s,t}$ is the observed market clustering and $\langle M_{s,t}\rangle^*$ is the expected value based on the benchmark model that we develop below. 
The observed value $M_{s,t}$ is divided by the expected value $\langle M_{s,t}\rangle^*$, so that deviations from the benchmark are scaled in terms of the expected value. The minimum value for the market clustering is minus one by definition and a market clustering of zero means that the market clustering has the same value as the expected value $\langle M_{s,t}\rangle^*$. We show detailed numerical examples in \Cref{fig:m_null,fig:m_example}.

The market clustering observation $M_{s,t}$ is defined as the number of shared securities, summed over all pairs of investors (which we visualize in \Cref{fig:m_obs}).  For each pair of firms, we first establish if they both trade in the security. If this is the case, we count the number of securities which these two firms are also trading simultaneously. The observed value of the market clustering $M_{s,t}$ for security $s$ during month $t$ is then given by:
\begin{equation}
M_{s,t} =
\sum\limits_{f}^{n_{F,t}}\sum\limits_{f'>f}^{n_{F,t}}
\Big(
a_{sf,t}a_{sf',t}
\sum\limits_{s'\neq s}^{n_{S,t}}a_{s'f,t}a_{s'f',t}\Big).
\label{eq:m_obs}
\end{equation}
The total number of firms and securities are denoted by $n_{F,t}$ and $n_{S,t}$, respectively. The summation $\sum\nolimits_{f}\sum\nolimits_{f'>f}$ runs over all possible pairs of investors and the summation $\sum\nolimits_{s' \neq s}$ runs, per pair of investors, over all securities except security $s$. The indicator $a_{sf,t}=1$ in case firm $f$ trades in security $s$ during month $t$ and $a_{sf,t}=0$ otherwise. $M_{s,t}$ measures all trading combinations within the pool of investors that trade in security $s$, forming a market clustering pattern or `motif'.\footnote{If investors in a security are otherwise not trading jointly, then $m_{s,t}=-1$ and we drop 3412 observations (5\%) of such cases since these observations are not relevant for our analysis.}

\begin{figure}[!htb]
    \centering
    \begin{adjustbox}{max width=\textwidth}
    \begin{tikzpicture}


    \node[draw=none] at (-0.5,11.5){a)};
    \node[draw=none] at (-0.5,9){b)};
    \node[draw=none] at (-0.5,4.5){c)};


    \def\a{10};\def\b{11.5};
    \foreach \x [count=\xi] in {2,2.5,...,6}
    	\node[draw,circle,inner sep =.5mm](s\xi) at (\x+3,\a){};
    \foreach \x [count=\xi] in {2,3,...,6}
    	\node[draw,circle,inner sep =.5mm](f\xi) at (\x+3,\b){};
    \node[draw=none](f) at (3,\b) {Firms};
    \node[draw=none](s) at (3,\a) {Securities};

\draw [decorate,decoration={brace,amplitude=10pt},xshift=-4pt,yshift=0pt] (5,\a+.1) -- (5,\b-.1) node [black,midway,xshift=-1cm] {\footnotesize Trades};

    \draw[draw,line width=.2mm] (s1) to (f2) to (s2) to (f3) to (s1);
    \draw[draw,line width=.2mm] (s1) to (f1) to (s3) to (f3) to (s1);
    \draw[draw,line width=.2mm] (s3) to (f1) to (s2) to (f3) to (s3);
    \draw[draw,line width=.2mm] (s7) to (f3) to (s9) to (f5) to (s7);
    \draw[draw,line width=.2mm] (s8) to (f4) to (s9) to (f5) to (s8);
    \draw[draw,line width=.2mm] (s4) to (f1);
    \draw[draw,line width=.2mm] (s5) to (f2);
    \draw[draw,line width=.2mm] (s6) to (f1);
    \draw[draw,line width=.2mm] (s6) to (f4);


\node[draw,circle,inner sep =.5mm,fill=red](red) at (5,10){};

    \def\a{7.5};\def\b{9};
    \foreach \x [count=\xi] in {2,2.5,...,6}
    	\node[draw,circle,inner sep =.5mm](s\xi) at (\x,\a){};
    \foreach \x [count=\xi] in {2,3,...,6}
    	\node[draw,circle,inner sep =.5mm](f\xi) at (\x,\b){};

    \draw[draw,line width=.2mm,dotted] (s1) to (f2) to (s2) to (f3) to (s1);
    \draw[draw,line width=.2mm,dotted] (s1) to (f1) to (s3) to (f3) to (s1);
    \draw[draw,line width=.2mm,dotted] (s3) to (f1) to (s2) to (f3) to (s3);
    \draw[draw,line width=.2mm,dotted] (s7) to (f3) to (s9) to (f5) to (s7);
    \draw[draw,line width=.2mm,dotted] (s8) to (f4) to (s9) to (f5) to (s8);
    \draw[draw,line width=.2mm,dotted] (s4) to (f1);
    \draw[draw,line width=.2mm,dotted] (s5) to (f2);
    \draw[draw,line width=.2mm,dotted] (s6) to (f1);
    \draw[draw,line width=.2mm,dotted] (s6) to (f4);

    \draw[draw,line width=.4mm] (s1) to (f1) to (s2) to (f2) to (s1);
    \node[draw,circle,inner sep =.7mm,fill=black] at (s1){};

    \foreach \x [count=\xi] in {2,2.5,...,6}
    	\node[draw,circle,inner sep =.5mm](s\xi) at (\x+6,\a){};
    \foreach \x [count=\xi] in {2,3,...,6}
    	\node[draw,circle,inner sep =.5mm](f\xi) at (\x+6,\b){};

    \draw[draw,line width=.2mm,dotted] (s1) to (f2) to (s2) to (f3) to (s1);
    \draw[draw,line width=.2mm,dotted] (s1) to (f1) to (s3) to (f3) to (s1);
    \draw[draw,line width=.2mm,dotted] (s3) to (f1) to (s2) to (f3) to (s3);
    \draw[draw,line width=.2mm,dotted] (s7) to (f3) to (s9) to (f5) to (s7);
    \draw[draw,line width=.2mm,dotted] (s8) to (f4) to (s9) to (f5) to (s8);
    \draw[draw,line width=.2mm,dotted] (s4) to (f1);
    \draw[draw,line width=.2mm,dotted] (s5) to (f2);
    \draw[draw,line width=.2mm,dotted] (s6) to (f1);
    \draw[draw,line width=.2mm,dotted] (s6) to (f4);

    \draw[draw,line width=.4mm] (s1) to (f1) to (s2) to (f3) to (s1);
    \node[draw,circle,inner sep =.7mm,fill=black] at (s1){};

    \def\a{5.5};\def\b{7};
    \foreach \x [count=\xi] in {2,2.5,...,6}
    	\node[draw,circle,inner sep =.5mm](s\xi) at (\x,\a){};
    \foreach \x [count=\xi] in {2,3,...,6}
    	\node[draw,circle,inner sep =.5mm](f\xi) at (\x,\b){};

    \draw[draw,line width=.2mm,dotted] (s1) to (f2) to (s2) to (f3) to (s1);
    \draw[draw,line width=.2mm,dotted] (s1) to (f1) to (s3) to (f3) to (s1);
    \draw[draw,line width=.2mm,dotted] (s3) to (f1) to (s2) to (f3) to (s3);
    \draw[draw,line width=.2mm,dotted] (s7) to (f3) to (s9) to (f5) to (s7);
    \draw[draw,line width=.2mm,dotted] (s8) to (f4) to (s9) to (f5) to (s8);
    \draw[draw,line width=.2mm,dotted] (s4) to (f1);
    \draw[draw,line width=.2mm,dotted] (s5) to (f2);
    \draw[draw,line width=.2mm,dotted] (s6) to (f1);
    \draw[draw,line width=.2mm,dotted] (s6) to (f4);

    \draw[draw,line width=.4mm] (s1) to (f1) to (s3) to (f3) to (s1);
    \node[draw,circle,inner sep =.7mm,fill=black] at (s1){};


    \foreach \x [count=\xi] in {2,2.5,...,6}
    	\node[draw,circle,inner sep =.5mm](s\xi) at (\x+6,\a){};
    \foreach \x [count=\xi] in {2,3,...,6}
    	\node[draw,circle,inner sep =.5mm](f\xi) at (\x+6,\b){};

    \draw[draw,line width=.2mm,dotted] (s1) to (f2) to (s2) to (f3) to (s1);
    \draw[draw,line width=.2mm,dotted] (s1) to (f1) to (s3) to (f3) to (s1);
    \draw[draw,line width=.2mm,dotted] (s3) to (f1) to (s2) to (f3) to (s3);
    \draw[draw,line width=.2mm,dotted] (s7) to (f3) to (s9) to (f5) to (s7);
    \draw[draw,line width=.2mm,dotted] (s8) to (f4) to (s9) to (f5) to (s8);
    \draw[draw,line width=.2mm,dotted] (s4) to (f1);
    \draw[draw,line width=.2mm,dotted] (s5) to (f2);
    \draw[draw,line width=.2mm,dotted] (s6) to (f1);
    \draw[draw,line width=.2mm,dotted] (s6) to (f4);

    \draw[draw,line width=.4mm] (s1) to (f2) to (s2) to (f3) to (s1);
    \node[draw,circle,inner sep =.7mm,fill=black] at (s1){};

    \def\a{3};\def\b{4.5};
    \foreach \x [count=\xi] in {2,2.5,...,6}
    	\node[draw,circle,inner sep =.5mm](s\xi) at (\x+3,\a){};
    \foreach \x [count=\xi] in {2,3,...,6}
    	\node[draw,circle,inner sep =.5mm](f\xi) at (\x+3,\b){};
    \node[draw=none](f) at (3,\b) {Firms};
    \node[draw=none](s) at (3,\a) {Securities};
    \draw [decorate,decoration={brace,amplitude=10pt},xshift=-4pt,yshift=0pt] (5,\a+.1) -- (5,\b-.1) node [black,midway,xshift=-1cm] {\footnotesize Trades};

    \draw[draw,line width=.2mm] (s1) to (f2) to (s2) to (f3) to (s1);
    \draw[draw,line width=.2mm] (s1) to (f1) to (s3) to (f3) to (s1);
    \draw[draw,line width=.2mm] (s3) to (f1) to (s2) to (f3) to (s3);
    \draw[draw,line width=.2mm] (s7) to (f3) to (s9) to (f5) to (s7);
    \draw[draw,line width=.2mm] (s8) to (f4) to (s9) to (f5) to (s8);
    \draw[draw,line width=.2mm,dotted] (s4) to (f1);
    \draw[draw,line width=.2mm,dotted] (s5) to (f2);
    \draw[draw,line width=.2mm,dotted] (s6) to (f1);
    \draw[draw,line width=.2mm,dotted] (s6) to (f4);

    \foreach \x [count=\xi] in {{\color{red} 4},4,2,0,0,0,1,1,2}
    	\node[draw=none] at (\xi/2+1.5+3,\a-0.5){\x};
    \node[draw=none] at (3,\a-0.5){$M_{i,t}$};
\end{tikzpicture} 
    \end{adjustbox}
    \caption{Example of the calculation of the observed market clustering $M_{s,t}$. \\
     {\footnotesize Panel a) A hypothetical bipartite trading network. Each line represents a buy or sell transaction. Panel b) Counting the market clustering motifs for the first security. In these four cases shown the trading pattern exist and therefore the first security has score four. This calculation is repeated for all other securities. The summation in \Cref{eq:m_obs} runs over all possibilities. Panel c) The same hypothetical trading situation with the observed market clustering $M_{s,t}$ for each security (lines that do not contribute to the market clustering measurements for any security are dotted).} }
    \label{fig:m_obs}
\end{figure}


We calculate the expected value of the market clustering based on the probability distribution $P(X|D_{t,\text{obs}})$, such that the expected value of the market clustering is only based on the observed degree sequences $D_{t,\text{obs}}$ (see \Cref{appendix:networkreconstruction}). The expected value of the market clustering is calculated as the sum over all configurations weighted by the probabilities:
\begin{align}
\langle M_{s,t}\rangle^*
&=
\sum\limits_{X\in\mathcal{G}}
P(X|D_{t,\text{obs}})M_{s}(X)
\\&=
\sum\limits_{X\in\mathcal{G}}
\sum\limits_f^{n_{F,t}-1}
\sum\limits_{f'>f}^{n_{F,t}}
\sum\limits_{s'\neq s}^{n_{S,t}}\Big[
 a_{sf}(X)P(a_{sf}(X)|D_{t,\text{obs}})
\nonumber\\& \hspace{4cm} \times a_{sf'}(X)P(a_{sf'}(X)|D_{t,\text{obs}})
\nonumber\\& \hspace{4cm} \times a_{s'f}(X)P(a_{s'f}(X)|D_{t,\text{obs}})
\nonumber\\& \hspace{4cm} \times a_{s'f'}(X)P(a_{s'f'}(X)|D_{t,\text{obs}})\Big]
\\&=
\sum\limits_f^{n_{F,t}-1}
\sum\limits_{f'>f}^{n_{F,t}}
\sum\limits_{s'\neq s}^{n_{S,t}}\Big[
P(a_{sf}=1|D_{t,\text{obs}})
P(a_{sf'}=1|D_{t,\text{obs}})
\nonumber\\&\hspace{4cm}\times
P(a_{s'f}=1|D_{t,\text{obs}})
P(a_{s'f'}=1|D_{t,\text{obs}})\Big]
\\&=
\sum\limits_{f}^{n_{F,t}-1}
\sum\limits_{f'>f}^{n_{F,t}}
\Big(
p_{sf,t}p_{sf',t}
\sum\limits_{s'\neq s}^{n_{S,t}}
p_{s'f,t}p_{s'f',t}\Big),
\label{eq:m_null}
\end{align}
where the last line only simplifies the notation. \Cref{fig:m_null} explains the summation process graphically.

\begin{figure}[!htb]
    \centering
    \resizebox*{\textwidth}{!}{
    \begin{tikzpicture}
    \node[draw=none] at (-0.5,14){a)};
    \node[draw=none] at (-0.5,11.7){b)};
    \node[draw=none] at (-0.5,9.7){c)};
    \node[draw=none] at (-0.5,7.7){d)};
    \node[draw=none] at (-0.5,5.2){e)};

    \def\a{12.5};\def\b{14};
    \foreach \x [count=\xi] in {2,2.5,...,6}
    	\node[draw,circle,inner sep =.5mm](s\xi) at (\x+3,\a){};
    \foreach \x [count=\xi] in {2,3,...,6}
    	\node[draw,circle,inner sep =.5mm](f\xi) at (\x+3,\b){};
    \node[draw=none](f) at (3,\b) {Firms};
    \node[draw=none](s) at (3,\a) {Securities};
    \draw [decorate,decoration={brace,amplitude=10pt},xshift=-4pt,yshift=0pt] (5,\a+.1) -- (5,\b-.1) node [black,midway,xshift=-1cm] {\footnotesize Trades};

    \draw[draw,line width=.2mm] (s1) to (f2) to (s2) to (f3) to (s1);
    \draw[draw,line width=.2mm] (s1) to (f1) to (s3) to (f3) to (s1);
    \draw[draw,line width=.2mm] (s3) to (f1) to (s2) to (f3) to (s3);
    \draw[draw,line width=.2mm] (s7) to (f3) to (s9) to (f5) to (s7);
    \draw[draw,line width=.2mm] (s8) to (f4) to (s9) to (f5) to (s8);
    \draw[draw,line width=.2mm] (s4) to (f1);
    \draw[draw,line width=.2mm] (s5) to (f2);
    \draw[draw,line width=.2mm] (s6) to (f1);
    \draw[draw,line width=.2mm] (s6) to (f4);

    \def\a{10.5};\def\b{11.7};
    \foreach \x [count=\xi] in {2,2.5,...,6}
    	\node[draw,circle,inner sep =.5mm](s\xi) at (\x+3,\a){};
    \foreach \x [count=\xi] in {2,3,...,6}
    	\node[draw,circle,inner sep =.5mm](f\xi) at (\x+3,\b){};

    \foreach \x [count=\xi] in {3,3,2,1,1,2,2,2,3}
    	\node[draw=none] at (\xi/2+1.5+3,\a+0.3){\x};
    \node[draw=none] at (3,\a+0.3){Security degree $d_{s,t}$};
    \foreach \x [count=\xi] in {5,3,5,3,3}
    	\node[draw=none] at (\xi+4,\b-0.3){\x};
    \node[draw=none] at (3,\b-0.3){Firm degree $d_{f,t}$};


    \def\a{8.2};\def\b{9.7};
    \foreach \x [count=\xi] in {2,2.5,...,6}
    	\node[draw,circle,inner sep =.5mm](s\xi) at (\x+3,\a){};
    \foreach \x [count=\xi] in {2,3,...,6}
    	\node[draw,circle,inner sep =.5mm](f\xi) at (\x+3,\b){};

    \foreach \x in {1,2,...,5}
    	\foreach \y in {1,2,...,9}
    {\draw[draw,line width=.2mm,dotted] (f\x) to (s\y);}
    \draw[draw,line width=.8mm,dotted] (f3) to (s1);		
    \node[draw=none,rectangle,fill=white] at (7,\a/2+\b/2) {$p_{13,t}=0.74$};
    \node[draw=none] at (3,\b/2+\a/2){Probabilities $p_{sf,t}$};

    \def\a{6.2};\def\b{7.7};
    \foreach \x [count=\xi] in {2,2.5,...,6}
    	\node[draw,circle,inner sep =.5mm](s\xi) at (\x,\a){};
    \foreach \x [count=\xi] in {2,3,...,6}
    	\node[draw,circle,inner sep =.5mm](f\xi) at (\x,\b){};
    \foreach \x in {1,2,...,5}
    	\foreach \y in {1,2,...,9}
    {\draw[draw,line width=.2mm,dotted] (f\x) to (s\y);}\draw[draw,line width=.8mm,dotted] (s1) to (f1) to (s2) to (f2) to (s1);
    \node[draw,circle,inner sep =.7mm,fill=black] at (s1){};
    \node[draw=none,rectangle,fill=white] at (4.5,\a/2+\b/2) {$p_{11,t}p_{12,t}p_{21,t}p_{22,t}=0.13$};

    \foreach \x [count=\xi] in {2,2.5,...,6}
    	\node[draw,circle,inner sep =.5mm](s\xi) at (\x+6,\a){};
    \foreach \x [count=\xi] in {2,3,...,6}
    	\node[draw,circle,inner sep =.5mm](f\xi) at (\x+6,\b){};
    \foreach \x in {1,2,...,5}
    	\foreach \y in {1,2,...,9}
    {\draw[draw,line width=.2mm,dotted] (f\x) to (s\y);}\draw[draw,line width=.8mm,dotted] (s1) to (f1) to (s3) to (f2) to (s1);
    \node[draw,circle,inner sep =.7mm,fill=black] at (s1){};
    \node[draw=none,rectangle,fill=white] at (10.5,\a/2+\b/2) {$p_{11,t}p_{12,t}p_{31,t}p_{32,t}=0.06$};

    \def\a{4.5};\def\b{5.4};
    \def\c{0.2};
    \foreach \x [count=\xi] in {0.4,1.3,...,7.7}
    	\node[draw,circle,inner sep =.5mm](s\xi) at (\x+3,\a){};
    \foreach \x [count=\xi] in {0.4,2.2,...,7.7}
    	\node[draw,circle,inner sep =.5mm](f\xi) at (\x+3,\b){};
    \node[draw=none](f) at (1.4,\b) {Firms};
    \node[draw=none](s) at (1.4,\a) {Securities};

    \draw[draw,line width=.2mm,dotted] (s1) to (f2) to (s2) to (f3) to (s1);
    \draw[draw,line width=.2mm,dotted] (s1) to (f1) to (s3) to (f3) to (s1);
    \draw[draw,line width=.2mm,dotted] (s3) to (f1) to (s2) to (f3) to (s3);
    \draw[draw,line width=.2mm,dotted] (s7) to (f3) to (s9) to (f5) to (s7);
    \draw[draw,line width=.2mm,dotted] (s8) to (f4) to (s9) to (f5) to (s8);
    \draw[draw,line width=.2mm,dotted] (s4) to (f1);
    \draw[draw,line width=.2mm,dotted] (s5) to (f2);
    \draw[draw,line width=.2mm,dotted] (s6) to (f1);
    \draw[draw,line width=.2mm,dotted] (s6) to (f4);

    \def\d{1.9}
    \foreach \x [count=\xi] in {5.3,5.3,2.8,0.8,0.9,2.8,2.8,2.8,5.3}
    	\node[draw=none] at (\xi*0.9+1.5+1,\a-\d+1.2){\x};
    \node[draw=none] at (1.4,\a-\d+1.2){$\langle M_{s,t}\rangle^*$};

    \foreach \x [count=\xi] in {4,4,2,0,0,0,1,1,2}
    	\node[draw=none] at (\xi*0.9+1.5+1,\a-\d+0.7){\x};
    \node[draw=none] at (1.4,\a-\d+0.7){$M_{s,t}$ };
    \foreach \x [count=\xi] in {-0.24,-0.24,-0.28,-1,-1,-1,-0.64,-0.64,-0.62}
    	\node[draw=none] at (\xi*0.9+1.5+1,\a-\d){\x};
    \node[draw=none] at (1.4,\a-\d){$m_{s,t}$ (\Cref{eq:m})};

    \draw[draw] (0.2,\a-\d+1.5) to (11.5,\a-\d+1.5);
    \draw[draw] (0.2,\a-\d+0.4) to (11.5,\a-\d+0.4);
    \draw[draw] (0.2,\a-\d-0.3) to (11.5,\a-\d-0.3);
\end{tikzpicture} 
    }
    \caption{Calculation of the benchmark model for the market clustering $\langle M_{s,t}\rangle^*$. \\
    {\footnotesize Panel a) The same hypothetical trading situation as in \Cref{fig:m_obs}. Panel b) The trading information is reduced to the degree sequences: The number of traded securities per firm and the number of trading firms per security. Panel c) The degree sequences are translated into a probability $p_{sf,t}$ for each firm-security pair (i.e. the probability of firm $f$ trading in security $s$ in month $t$). The probability and degree sequences hold the same information, since the expected value of the number of connections for each node equals the degree. Panel d) The probability of occurrence of the market clustering motifs equals the product of the four probabilities between the four involved nodes (see \Cref{eq:m_null}). The expected value of the market clustering per security is the sum of all probabilities for motifs that are connected to the security. The first two motifs for the first security are shown. This calculation is repeated for each security. Panel e) The benchmark model market clustering $\langle M_{s,t}\rangle^*$ for each security, the observed market clustering from \Cref{fig:m_obs}, and the final market clustering measures, according to \Cref{eq:m}, respectively.}}
    \label{fig:m_null}
\end{figure}

\clearpage

The market clustering $m_{s,t}$ measures the degree of clustering for security $s$ among its traders.
\Cref{fig:m_example} shows examples of the performance of the method in two hypothetical situations. Firstly, the model assigns a lower value to securities which are involved in multiple clusters. Arguably, the involvement in multiple clusters enhances the diversity of the investors group and would probably stabilize the price dynamics. Secondly, \Cref{fig:m_example} shows that the model is able to indicate to what extent the security is involved in the cluster. Homogeneous trading behaviour is indicated by a relatively high percentage of overlapping trades. Therefore, the number of trades that do not overlap must lower the market clustering measure. This condition is satisfied as can be seen in the second example in \Cref{fig:m_example}.

\begin{figure}[!htb]
    \centering
    \begin{adjustbox}{max width=\textwidth}
    \begin{tikzpicture}
    \def\a{4};
    \def\b{5.5};

    \node[draw=none,anchor=north west,text width = 4cm] at (-6,\b) {Example 1: Involvement in multiple clusters lowers $m_{s,t}$.};

    \foreach \x [count=\xi] in {2,3,...,6}
    	\node[draw,circle,inner sep=1mm](s\xi) at (\x,\a){};
    \foreach \x [count=\xi] in {2,3,...,6}
    	\node[draw,circle,inner sep=1mm](f\xi) at (\x,\b){};
    \node[draw=none](f) at (0,\b) {Firms};
    \node[draw=none](s) at (0,\a) {Securities};
    \draw[-,line width=0.5mm] (f) to [out=-90,in=90] node[draw=none,rectangle,fill=white,anchor=center]{Trades} (s);

    \draw[draw,line width=.5mm] (f1) to (s1) to (f2);
    \draw[draw,line width=.5mm] (f1) to (s2) to (f2);
    \draw[draw,line width=.5mm] (f1) to (s3) to (f2);
    \draw[draw,line width=.5mm] (s3) to (f4);
    \draw[draw,line width=.5mm] (f4) to (s3) to (f5);
    \draw[draw,line width=.5mm] (f4) to (s4) to (f5);
    \draw[draw,line width=.5mm] (f4) to (s5) to (f5);

    \def\d{\a-2}
    \foreach \x [count=\xi] in {2.6,2.6,6,2.6,2.6}
    	\node[draw=none] at (\xi+1,\d+1.2){\x};
    \node[draw=none] at (0,\d+1.2){$\langle M_{s,t}\rangle^*$ (\Cref{fig:m_null})};

    \foreach \x [count=\xi] in {2,2,4,2,2}
    	\node[draw=none] at (\xi+1,\d+0.7){\x};
    \node[draw=none] at (0,\d+0.7){$M_{s,t}$(\Cref{fig:m_obs})};

    \foreach \x [count=\xi] in {-0.24,-0.24,-0.33,-0.24,-0.24}
    	\node[draw=none] at (\xi+1,\d){\x};
    \node[draw=none] at (0,\d){$m_{s,t}$ (\Cref{eq:m})};

    \draw[draw] (-1.5,\d+1.5) to (7,\d+1.5);
    \draw[draw] (-1.5,\d+0.4) to (7,\d+0.4);
    \draw[draw] (-1.5,\d-0.3) to (7,\d-0.3);

    \def\a{-1};
    \def\b{0.5};

    \node[draw=none,anchor=north west,text width = 4cm] at (-6,\b) {Example 2: Market clustering $m_{s,t}$ indicates the degree of involvement in the cluster.};

    \foreach \x [count=\xi] in {2,3,...,6}
    	\node[draw,circle,inner sep=1mm](s\xi) at (\x,\a){};
    \foreach \x [count=\xi] in {2,3,...,6}
    	\node[draw,circle,inner sep=1mm](f\xi) at (\x,\b){};
    \node[draw=none](f) at (0,\b) {Firms};
    \node[draw=none](s) at (0,\a) {Securities};
    \draw[-,line width=0.5mm] (f) to [out=-90,in=90] node[draw=none,rectangle,fill=white,anchor=center]{Trades} (s);

    \draw[draw,line width=.5mm] (s1) to (f1) to (s2);
    \draw[draw,line width=.5mm] (s1) to (f2) to (s2);
    \draw[draw,line width=.5mm] (s1) to (f3) to (s2);
    \draw[draw,line width=.5mm] (f1) to (s3) to (f3);
    \draw[draw,line width=.5mm] (s1) to (f2) to (s4);
    \draw[draw,line width=.5mm] (s3) to (f1) to (s4);
    \draw[draw,line width=.5mm] (s5) to (f4) to (s4);
    \draw[draw,line width=.5mm] (f3) to (s5) to (f5);
    \draw[draw,line width=.5mm] (f3) to (s3) to (f2);

    \def\d{\a-2}
    \foreach \x [count=\xi] in {6.5,6.5,6.5,6.5,6.5}
    	\node[draw=none] at (\xi+1,\d+1.2){\x};
    \node[draw=none] at (0,\d+1.2){$\langle M_{s,t}\rangle^*$ (\Cref{fig:m_null})};

    \foreach \x [count=\xi] in {7,7,7,3,0}
    	\node[draw=none] at (\xi+1,\d+0.7){\x};
    \node[draw=none] at (0,\d+0.7){$M_{s,t}$ (\Cref{fig:m_obs})};

    \foreach \x [count=\xi] in {0.08,0.08,0.08,-0.54,-1}
    	\node[draw=none] at (\xi+1,\d){\x};
    \node[draw=none] at (0,\d){$m_{s,t}$ (\Cref{eq:m})};

    \draw[draw] (-1.5,\d+1.5) to (7,\d+1.5);
    \draw[draw] (-1.5,\d+0.4) to (7,\d+0.4);
    \draw[draw] (-1.5,\d-0.3) to (7,\d-0.3);
\end{tikzpicture}
    \end{adjustbox}
    \caption{Example of the computation of $m_{s,t}$. \\ 
    {\footnotesize Example 1. The market clustering is lower when a security is involved in multiple clusters at once. In this configuration two clusters exist, one on the left and one on the right. The security in the middle is involved in both clusters. The final market clustering is lower for the security in the middle, because its four connected firms are not mutually clustered. Example 2. The market clustering $m_{s,t}$ indicates the involvement in the market cluster. All securities are traded by three firms each. The left three firms are almost fully clustered. The final market clustering $m_{s,t}$ indicates to which extent the securities are involved in the cluster.} }
    \label{fig:m_example}
\end{figure}

\clearpage

\subsection{Measuring price instability}\label{subsection:priceinst}

We measure stock price instability with statistics that focus on tail behavior of the stock return distribution. We analyze the skewness, the kurtosis, the tail indices, the number of outliers, and the changes in the left and right $5\%$ quantiles. The latter two can also be interpreted as changes in downside risk and upward potential and are better manageable on the time-series dimension. \cite{ang_downside_2006} show that sensitivities to downside market movements are priced in addition to the common risk factors. Thus if market clustering leads to changes in downside risk, it implicitly shows up in the price dynamics.

Skewness and kurtosis are measures of the shape of the complete log return distribution while the outlier count and the tail index are focused on the tails of the distributions -- the extreme returns. The tail index (i.e, Hill's estimator) measures the fatness of the tail according to the power law distribution. We count the number of outliers by sequentially applying the generalized Grubbs' test until no outliers are detected. The skewness, Hill indices and outlier count also allow us to distinguish the effect on price instability for up- and downward shocks separately. We measure the size of the price fluctuation relative to the yearly standard deviation of the stocks, i.e., we divide the log returns by the yearly standard deviation per stock. Complementary to the volatility normalization, we investigate the influence of market clustering on the variance and the Median Average Deviation (MAD), which is more robust to outliers than the variance.

Value-at-Risk (VaR) -- often used in risk management and regulation -- is an obvious choice for quantifying the downside risk. We focus on a single stock 5\% VaR obtained via historical bootstrap from daily returns. Historical simulation risk measures depend on the level of volatility in the sample. However, our quantile-based variables measure change over time and relative deviation from the cross-sectional median and, as such, they are not affected by volatility clustering. More precisely, for the monthly data set we define:
\[ \operatorname{VaR\_chg}_{st} = 100\Big( \frac{\operatorname {VaR}_s(t-11,t)}{\operatorname {VaR}_s(t-12,t-1)}-1 \Big),\]
\[ \operatorname{VaR\_dev}_{st} = 100\Big( \frac{\operatorname {VaR}_s(t-11,t)}{\operatorname{median}(\{ \operatorname {VaR}_s(t-11,t); s=1,2,\dots,N \})}-1 \Big), \]
where $\operatorname {VaR}_s(t_1,t_2)$ denotes a 5\% VaR for stock $s$ at the end of month $t_2$ obtained via historical bootstrap from daily prices over the period from month $t_1$ to month $t_2$.

Similarly, to capture tail asymmetries, we define changes in Value-at-Luck (VLuck):
\[ \operatorname{VLuck\_chg}_{st} = 100\Big( \frac{\operatorname {VLuck}_s(t-11,t)}{\operatorname {VLuck}_s(t-12,t-1)}-1 \Big),\]
where $\operatorname {VLuck}_s(t_1,t_2)$ denotes a 95\% VaR for stock $s$ at the end of month $t_2$. 

\subsection{Stochastic dominance and causality for groups}\label{subsection:group}

We now compare the distributions of the price instability measures for low and high market clustering. Firstly, the securities are ordered according to their market clustering measure. Secondly, the securities are divided into three groups:
the lowest ($L$) and the highest ($H$) 33\%. We ignore the middle group in the remainder.
  Finally, we collect all time series price instability measures per time window per group and assess first and second order stochastic dominance of the distributions for group $L$ and $H$.

We use three tests to indicate the differences between the distributions of groups $L$ and $H$. The Kolmogorov-Smirnov (KS) test and the Mann-Whitney-Wilcoxon (MWW) test are both nonparametric tests for unpaired samples. The $\chi^2$ test is used instead of the KS test in case of binned data, because the KS test is unreliable when the number of ties is high. The KS test is sensitive to any discrepancy in the cumulative distribution function and serves as a test for the first-order stochastic dominance. The MWW is mainly sensitive to changes in the median and aids to evaluate the second-order stochastic dominance. We use visual inspection of the cumulative distributions to study the nature of the discrepancies to interpret the test results.

Using difference-in-differences approach allows us to benefit from the partial coverage of our data et and dispel concerns over reversed causation. A concern could be that rather than market clustering causing price instability (null hypothesis), unstable and risky stocks might attract traders that prefer to trade in clusters of like-minded traders. In order to assess the effect of clustered trading in a mimicked experimental research setting, we construct a so-called control group from the stocks that are mainly traded by investors \emph{not} included in the our data set. We look at the relation between market clustering and kurtosis in the control group. A significant relation would be speak against causality. 

\FloatBarrier

\subsection{Dynamic panel data framework}\label{subsection:dynamicpanel}

\noindent The last part of the analysis applies a dynamic unbalanced panel data model. We aim to strengthen the high and low market clustering group results by exploring a) the possible drivers of the clustering measure and b) the effect that the clustering measure has on price instability. We tackle two questions in the model for the market clustering drivers. One, a group of investors may choose particular stocks because of their (latent) properties. We also include the properties that quantify a stock's riskiness and instability as an additional test on reverse causality, mentioned in the previous section. Two, crowded trading activity may depend on certain market conditions. We look at the effect of the perceived trend and volatility. In the second application we investigate the relation between changes in left and right quantiles of log returns distribution and clustering in individual stocks. In particular, we are interested to see whether a higher clustering measure leads to larger changes in downside risk VaR and upside potential VLuck after controlling for other possible individual stock risk determinants.

The general representation of the model with both lagged dependent and independent variables are included, possibly of different depth, is as follows:
\begin{equation} \label{eq:dynamicpanel}
y_{st} = \sum_r \rho_r y_{s,t-r} + \sum_p \beta_{p}^\text{T}x_{s,t-p}+\alpha_s+\epsilon_{st},r=1,2,\dots,p=0,1,2,\dots,
\end{equation}
where $y_{st}$ is a dependent variable, i.e., the clustering measure or the price instability measure depending on exact specification, $x_{s,t}$ is a vector of considered covariates, $\alpha_s$ is an individual effect, $rho_r$ and $\beta_{p}$ denote model parameters, $\epsilon_{st}$ is idiosyncratic error term, $ s=1,\dots,N$, and $t=1,\dots,T$.

We opt for the fixed effects model and treat $\alpha_s$ as a set of $N$ additional parameters. We do not employ time dummies for two reasons. First, time dummies would preclude including time-only varying variables of interest (like the market factor MKTF and market volatility VIX). Second, incorporation of time dummies is more suitable for panels with very small $T$ (in fact, most of our efforts to run the dynamic model with both fixed and time effects result in singularity issues). We estimate \Cref{eq:dynamicpanel} with the System GMM. In particular, a two-step estimator with \cite{windmeijer_finite_2005} correction for standard errors is used. Estimation is carried out with the \textsf{R} package \textsf{plm} \citep{croissant_panel_2008}.

Our methodology has several attractive properties. First, individual effects are allowed to be correlated with $x_{st}$ -- a likely case in our data as, e.g., firms in certain industries may have higher dividend yields or price-to-book ratios than others. Second, the fixed effects approach accounts for unobserved heterogeneity bias. All (practically) static cross-sectional stock features, like sector or exchange, are by default incorporated into $\alpha_s$ terms. Last but not least, we address potential endogeneity issue due to simultaneity. We hypothesize that an increase in the clustering measure leads to larger changes in downside risk. However, it is also possible that some stocks are more likely to end up in cluster trades because of their risk profile. We disentangle the causality by producing internal instruments for the right hand side variable CLUST that is not strictly exogenous.

A common approach is to use all possible lags and variables to construct GMM-style instruments. \cite{roodman_note_2009} warns that too many instruments result in model validity issues and, specifically, false estimation outcomes and low power of overidentification tests. Roodman suggests collapsing the instruments and using only certain lags to overcome the instrument proliferation. \cite{wintoki_endogeneity_2012} show that both collapsing the instruments and the size of cross-section increase the power of Sargan-Hansen-J test. However, these techniques do not guarantee elimination of the problem and thus we also perform sensitivity tests on the choice of instruments as prescribed by \cite{roodman_note_2009}.

\section{Data}
	\label{section:data}

The data have been collected as part of the Markets in Financial Instruments Directive (MiFID). MiFID is a European Union (EU) law to regulate investment services across the European Economic Area (EEA).\footnote{Directive 2004/39/EC. Official Journal of the European Union, 2004; amended subsequently: Directive 2008/10/EC. Official Journal of the European Union, 2008.} The directive applies to all firms that perform investment services and activities. Firms that only perform ancillary services are exempted. 'Post-trade transparency' is the key aspect of MiFID mandating the authorities to collect the data used here. The post-trade transparency regulation requires all firms to report all trades in all listed stocks, including the time, the price and number of units to the supervisory authorities immediately after the trade. MiFID only contains information about the transactions and thus holdings that are not traded are not in the data.

Although MiFID collects data on a EU level, Dutch authorities only have access to the transactions of Dutch banks and investment firms. In particular, the data cover the investments in financial instruments of 86 Dutch banks and investment firms. The time span of the data covers January 2009 through April 2015. The annual cross-sectional analysis (see \Cref{section:results}) is hence done for the period January 2009--December 2014.
Only the face-to-market firms report their transactions. The data contains trades by the reporter as principal trader and as agents. For the market as a whole, agent-trades form a limited part and are roughly at 10\% 
 of volume/trades. Furthermore, although we do not have information on the identity of the clients it is likely that they are non-financial firms or retail clients and hence will be very heterogeneous in their trading strategies. For the moment, we thus concentrate on trades entered into as principal. In case a principal trader performs transactions via a broker, only the broker reports the transaction, but we do see it in the data.

Contrary to portfolio holdings data sets, such as the ESCB Securities Holding Statistics, which show only shifts in the portfolio holdings, the MiFID data set contains all buy and sell transactions separately. We aggregate these transaction level data to a monthly frequency, split by the total number of buy and sell transactions. Aggregation of data is necessary because trading clusters do not emerge instantaneously, but rather over time. This choice facilitates our research design, meaning that we can derive price instability metrics from less noisy daily data instead of intra-day observations. 

To improve the comparability of the price dynamics, we perform the cross-sectional comparison only for equities and exclude bond trading. In general, the price dynamics and trading behaviour differ markedly between equity and bond markets. In contrast to equities, most bonds are not unique since bonds issued by the same entity but of different maturities are to a degree interchangeable (in case no arbitrage opportunities exist). In addition, we want to abstract from the dynamics at the beginning or end of the lifetime of a security (e.g., an IPO or a default). Thus we select 976 equities that are traded during each month in the period January 2009--April 2015.

The data source for the daily stock return time series is Bloomberg Professional. In case securities in our data are traded at multiple exchanges, Bloomberg chooses between the exchanges automatically. In case no transactions are registered during the day, the price of the security is kept at the price of the last transaction. After inspecting the price series for outliers, we remove two time series of penny stocks with excessive return volatility.

We apply a panel data framework for securities classified as common stocks in the Bloomberg database. The initial sample of $976$ equities contains $583$ common stocks. We remove $16$ stocks for which the average price does not exceed $1$ EUR, then $24$ stocks which are thinly traded (more than $10\%$ of days during the trading period without a single transaction), $2$ stocks with non-euro currency data, and $2$ stocks with suspiciously large values for some fundamentals. Next, we apply the turnover requirements for each year as in the first part of the analysis. Many of the stocks qualify for multiple years, in total we have $N=269$ unique stocks and $T=76$ months. The number of stocks across years fluctuate between 203 and 234.

The summary of explanatory variables and applied transformations is shown in \Cref{table:indep_vars}. We consider a wide variety of potential risk and trading behavior drivers: stock market conditions, individual stock performance, liquidity, and fundamentals.  MKTF and VIX are only time-varying variables, LEV3 monthly values repeat for the same fiscal quarter, and all other variables vary per stock per time period. Non-time-varying variables, like the sector of the issuer, cannot be explicitly accommodated in a panel framework with fixed effects.

\begin{table}[!htb]
\centering
	\begin{tabular}{lp{.8\textwidth}}
	\toprule
	Variable & Definition \\
	\midrule
		MKTF & Fama and French market factor for Europe, returns in \% for a month\\
		VIX & The CBOE Volatility index as a proxy to market conditions, level at the end of a month\\
		MOM & 12/6-month average of monthly returns in \% at the end of a month\\
		MCAP & Log of market capitalization in $10^6$ EUR at the end of a month\\
		ILLIQ & A daily ratio of absolute stock return to its euro volume, averaged over a month, also known as \cite{amihud_illiquidity_2002} liquidity measure, to reduce heteroskedasticity we transform as $\log(\text{ratio}+10^{-6})$ \\
		PB3 & Price-to-book ratio with a 3-month publication lag at the end of a month\\
		DY & 12-month trailing dividend yield, in \% at the end of a month, we set not available values to $0$ \\
		LEV3 & Ratio of long-term debt to capital with a 3 month publication lag at the end of a month\\
	\bottomrule
	\end{tabular}
\caption{Description of variables.}
\label{table:indep_vars}
\end{table}

The Fama and French market factor for Europe is downloaded from the Kenneth French library.\footnote{See \url{https://goo.gl/pZVmqe}.} The VIX index comes from Chicago Board Options Exchange website.\footnote{See \url{https://goo.gl/zMCTa}.} We obtain all stock specific information via Bloomberg terminal. 

\begin{table}[!htb]
\centering
\resizebox{\columnwidth}{!}{
\begin{tabular}{lrrrrrrrr}
  \toprule
 & N.Obs. & Mean & Median & St.dev. & Min & Max & Between & Within \\ 
  \midrule
  VaR\_dev & 16216 & 6.43 & 0.00 & 45.80 & -75.06 & 795.32 & 0.595 & 0.007 \\ 
  VaR\_chg & 16216 & -0.80 & 0.00 & 7.68 & -39.10 & 82.10 & 0.010 & 0.356 \\ 
  VLuck\_chg & 16216 & -0.50 & 0.00 & 7.69 & -59.74 & 98.78 & 0.011 & 0.190 \\ 
  CLUST & 15896 & 0.04 & 0.06 & 0.28 & -1.00 & 6.58 & 0.245 & 0.041 \\ 
  MKTF & 16216 & 1.09 & 1.02 & 5.97 & -12.33 & 13.86 & 0.003 & 1.000 \\ 
  VIX & 16216 & 21.09 & 18.38 & 8.12 & 11.40 & 46.35 & 0.052 & 1.000 \\ 
  MOM & 16204 & 0.64 & 0.71 & 3.48 & -18.88 & 27.71 & 0.106 & 0.338 \\ 
  MCAP & 16216 & 7.06 & 6.99 & 2.15 & 0.93 & 12.21 & 0.970 & 0.004 \\ 
  ILLIQ & 16051 & -3.66 & -4.40 & 4.47 & -13.82 & 12.48 & 0.957 & 0.007 \\ 
  PB3 & 15025 & 1.95 & 1.35 & 2.50 & 0.05 & 71.67 & 0.470 & 0.010 \\ 
  DY & 16216 & 2.88 & 2.15 & 4.53 & 0.00 & 157.78 & 0.303 & 0.032 \\ 
  LEV3 & 15916 & 26.39 & 25.35 & 20.42 & 0.00 & 159.01 & 0.813 & 0.007 \\ 
   \bottomrule
\end{tabular}

\caption{Descriptive statistics. \\ {\footnotesize For each variable the table presents the number of available observations, mean, median, standard deviation, minimum, maximum, and the proportions of between and within variation. Note that the proportions of variation do not add up to one because the panel is unbalanced.}}
\label{table:desc_stats}
}
\end{table}

\Cref{table:desc_stats} presents the descriptive statistics for variables in the panel data models. We discuss the last two columns as they offer the most valuable insights with regard to methodological choices. VaR\_chg, the percentage point change in the VaR, has substantial within variation of $0.356$ and thus the fixed effects model seems suitable for it. VLuck\_chg, the percentage point change in the upside VaR, has somewhat smaller yet acceptable within variation. VaR\_dev, the deviation from the period median, by definition varies mostly across groups (variation of $0.595$) and much less over time (variation of $0.007$), thus we will use it as a possible driver of clustered trading, but not as a dependent variable. We can expect effects of MKTF, VIX, and MOM to be estimated precisely because of (relatively) high within variation proportions of $1.000$, $1.000$, and $0.338$. Perhaps we will see effects of CLUST and DY as well, but the rest of the variables are likely to have high standard errors. MCAP and ILLIQ have such high between groups variation (close to one) that their explanatory power may be subsumed by fixed effects. 

\begin{table}[!htb]
\centering
\resizebox{\columnwidth}{!}{
\begin{tabular}{lrrrrrrrrrrrrrr}
  \toprule
	&	VaR\_dev	&	VaR\_chg	&	VLuck\_chg	&	CLUST	&	MKTF	&	VIX	&	MOM	&	MCAP	&	ILLIQ	&	PB3	&	DY	&	LEV3	\\
\midrule																									
VaR\_dev	&		&	0.073	&	0.017	&	-0.061	&	-0.002	&	-0.047	&	-0.022	&	-0.268	&	0.169	&	-0.067	&	-0.178	&	0.010	\\
VaR\_chg	&	0.000	&		&	0.000	&	0.018	&	-0.203	&	0.186	&	-0.204	&	-0.039	&	0.047	&	-0.006	&	0.063	&	-0.009	\\
VLuck\_chg	&	0.031	&	0.293	&		&	0.020	&	-0.098	&	0.154	&	-0.253	&	-0.053	&	0.047	&	-0.010	&	0.081	&	-0.005	\\
CLUST	&	0.000	&	0.020	&	0.011	&		&	0.004	&	0.102	&	-0.036	&	-0.041	&	0.058	&	-0.024	&	0.031	&	-0.025	\\
MKTF	&	0.847	&	0.000	&	0.000	&	0.613	&		&	-0.245	&	0.026	&	0.015	&	-0.018	&	-0.018	&	-0.021	&	0.015	\\
VIX	&	0.000	&	0.000	&	0.000	&	0.000	&	0.000	&		&	-0.315	&	-0.004	&	0.033	&	-0.042	&	0.127	&	0.054	\\
MOM	&	0.005	&	0.000	&	0.000	&	0.000	&	0.001	&	0.000	&		&	0.091	&	-0.088	&	0.152	&	-0.202	&	-0.023	\\
MCAP	&	0.000	&	0.000	&	0.000	&	0.000	&	0.050	&	0.612	&	0.000	&		&	-0.620	&	0.046	&	0.096	&	0.227	\\
ILLIQ	&	0.000	&	0.000	&	0.000	&	0.000	&	0.027	&	0.000	&	0.000	&	0.000	&		&	-0.069	&	-0.036	&	-0.165	\\
PB3	&	0.000	&	0.468	&	0.208	&	0.003	&	0.030	&	0.000	&	0.000	&	0.000	&	0.000	&		&	-0.039	&	0.015	\\
DY	&	0.000	&	0.000	&	0.000	&	0.000	&	0.007	&	0.000	&	0.000	&	0.000	&	0.000	&	0.000	&		&	0.059	\\
LEV3	&	0.191	&	0.276	&	0.501	&	0.002	&	0.062	&	0.000	&	0.003	&	0.000	&	0.000	&	0.065	&	0.000	&		\\

   \bottomrule
\end{tabular}

\caption{Correlations of pooled variables. \\ {\footnotesize The table presents Pearson correlations of pooled variables in the upper right triangles and the corresponding p-values to test for zero coefficient in the lower left triangles.}}
\label{table:corrs}
}
\end{table}

\Cref{table:corrs} provides information about comovements of variables. The relative level of the downside risk, VaR\_dev, has a significant negative correlation with CLUST of $-0.061$. The changes in downside risk and upside potential, i.e., VaR\_chg and VLuck\_chg, have marginally significant positive correlations with CLUST of $0.018$ and $0.020$. The negative correlation with the risk level is somewhat puzzling. Possibly, there is some more complicated interplay between the two than just a linear relationship with one-way temporal causality. Therefore, we will consider dynamic models in the presence of endogeneity.

The direction of movements also differs between VIX and the risk-based price instability measures. VIX is positively related to VaR\_chg and VLuck\_chg, but is negatively related to VaR\_dev. The latter can perhaps be explained by reasoning that when the market conditions are bad, the riskiness of all stocks becomes more similar and thus the deviations from the median decrease.

MKTF seems unrelated to VaR\_dev ($p-\text{value}=0.847$), but strongly negatively related to VaR\_chg with a correlation coefficient of $-0.203$ (expected as higher market returns reduce downside risk). MKTF is negatively related to VLuck\_chg with a significant correlation coefficient of $-0.098$. The cross-sectionally dispersed variables MCAP and ILLIQ have stronger correlations with the level of risk than the change in risk, supporting the findings for between and within groups variation. MOM, a variable with evident time-variation, shows ten-fold greater negative correlations with risk and gain changes over time than the level of risk. Further, the correlations of all independent variables are within reasonable magnitudes and thus multicolinearity should not be a problem.

\begin{table}[!htb]
\centering
\begin{tabular}{crrrrrrrr}
  \toprule
  \multirow{2}{*}{Lag} & \multicolumn{2}{c}{VaR\_dev} & \multicolumn{2}{c}{VaR\_chg} & \multicolumn{2}{c}{VLuck\_chg} & \multicolumn{2}{c}{CLUST} \\ \cmidrule(r){2-3} \cmidrule(r){4-5} \cmidrule(r){6-7} \cmidrule(r){8-9}
 & ACF & PACF & ACF & PACF & ACF & PACF  & ACF & PACF\\ 
  \midrule
1	&	93.31	&	93.31	&	43.49	&	43.49	&	35.69	&	35.69	&	38.29	&	38.29	\\
2	&	85.87	&	1.12	&	17.10	&	5.20	&	18.59	&	5.58	&	27.14	&	11.52	\\
3	&	83.64	&	1.86	&	5.20	&	2.97	&	8.18	&	2.60	&	24.16	&	10.41	\\
4	&	79.18	&	0.00	&	0.74	&	0.74	&	2.23	&	1.12	&	18.96	&	4.83	\\
5	&	73.61	&	0.00	&	0.37	&	1.12	&	1.12	&	0.74	&	12.64	&	2.60	\\
6	&	63.57	&	0.00	&	0.00	&	0.37	&	0.37	&	1.12	&	8.92	&	1.49	\\
7	&	53.53	&	0.00	&	1.12	&	0.74	&	0.00	&	0.74	&	5.95	&	1.12	\\

   \bottomrule
\end{tabular}

\caption{Percentage of significant ACFs and PACFs for VaR\_dev, VaR\_chg, VLuck\_chg, and CLUST.\\ {\footnotesize We obtain the autocorelations and partial autocorrelations for $269$ time series per variable in the panel data set. The table contains the percentages of cases with significant coefficients for the first seven lags.}}
\label{table:panel_acfs}
\end{table}

\FloatBarrier

The autocorrelations and partial autocorrelations (\Cref{table:panel_acfs}) indicate dynamic nature of all risk series and the clustering measure. VaR\_dev resembles an AR(1) process, again showing that it is not suitable as a dependent variable in the models. Two lags seem an appropriate starting point for the dynamic models explaining CLUST, VaR\_chg, and VLuck\_chg.

\section{Results }
	\label{section:results}

\subsection{Group comparison}\label{subsection:groupres}

We compare distributions of price instability measures between the buckets of stocks with high and low market clustering. Our first key observation is that there seems to be a relation between the kurtosis of the log return time series and market clustering. Panel a) in \Cref{table:stability} shows an overview of the results of the 24 test cases (MAD, variance, skewness, kurtosis where in each cell we show the results of the Kolmogorov-Smirnov (KS) test and the Mann-Whitney-Wilcoxon (MWW) test). For all six years both tests give significant indication for a positive relation between market clustering and the kurtosis (with significance level of 2.5\%). The test results are confirmed visually by the distance between the graphs of the cumulative kurtosis distribution for low and high market clustering (see \Cref{fig:cumulative_skew_kurt}). The cumulative distribution of the kurtosis in high market clustering group stochastically dominates the cumulative distribution of the kurtosis in low market clustering group. Since the sample kurtosis is a measure of tail extremity and peakedness, the stocks with a higher (lower) market clustering tend to have log return distributions which are more (less) peaked and have (less) fat tails.

\begin{table}[!htb]
\centering
\begin{tabular}{lrcccccc}
\hline
 & & 2009 & 2010 & 2011 & 2012 & 2013 & 2014 \\
\hline

a) & MAD & $==$ & $==$ & $==$ & $==$ & $==$ & $++$ \\
&Variance & $+=$ & $=+$ & $++$ & $==$ & $++$ & $++$ \\
&Skewness & $+=$ & $++$ & $++$ & $==$ & $++$ & $++$ \\
&Kurtosis & $++$ & $++$ & $++$ & $++$ & $++$ & $++$ \\

\hline
b) &Hill index neg. & $-=$ & $==$ & $=-$ & $==$ & $==$ & $==$ \\
&Hill index pos. & $--$ & $--$ & $--$ & $==$ & $--$ & $--$ \\

\hline
c) & Outliers neg. & $\neq$$=$ & $=$$=$ & $=$$=$ & $=$$=$ & $=$$=$ & $\neq$$=$ \\
&Outliers pos. & $\neq$$+$ & $\neq$$+$ & $\neq$$+$ & $=$$=$ & $\neq$$+$ & $\neq$$+$ \\

\hline
\end{tabular}

\caption{Testing for a relation between market clustering and price instability -- annual window. \\
{ \footnotesize
We compare the distributions of the four time series measures in Panel a), the Hill indices of the negative and positive tails in Panel b), and the number of outliers per time series in Panel c) over six years between two groups of stocks: the lowest 33\% and the highest 33\% of the stocks, ranked according to their market clustering measure. The table shows for each comparison two test results. In panels a and b, the first is the Kolmogorov-Smirnov (KS) test and the second is the Mann-Whitney-Wilcoxon (MWW) test. The critical value is 0.025 for both tests. A `+'/`-'/`=' sign means that the distribution for high market clustering exceeds/undercuts/is equal to the distribution for low market clustering.  In panel c the first is the $\chi ^2$-test (critical value: 0.05) and the second is the MWW test (critical value: 0.025). Contrary to the KS test, the $\chi^2$ test results indicates only whether the hypothesis of homogeneity is accepted ('$=$') or rejected ('$\neq$'). Note that we do not show 2015 because the comparison with other years would be difficult as we have significantly fewer observations.
} }
\label{table:stability}
\end{table}

The stochastic dominance of distributions of considered price instability measures conditional on positive and negative tail in high vs. low market clustering groups indicate that market clustering relates to a relatively heavier tail for the positive tail of the log return distribution and not for the negative tail. The results for the Hill indices (Panel b) in \Cref{table:stability} and \Cref{fig:cumulative_hill_out}) show that only the fatness of the positive tail relates to market clustering. Distribution of positive tail index in low clustering group dominates distribution of positive tail index in high clustering group. Here a lower index implies fatter tails. The results for the outlier count (Panel c) in \Cref{table:stability} and \Cref{fig:cumulative_hill_out}) also show a clear relation between the number of positive outliers and market clustering and not for the number of negative outliers. Distribution of positive outliers in high clustering group stochastically dominates distribution of positive outlier in low clustering group. The stochastic dominance of distributions of price instability measures for the negative tail cannot be established. The tests in \Cref{table:stability} provide evidence for neither first nor second order stochastic dominance.
 
The positive relation between the skewness and market clustering in \Cref{table:stability} and \Cref{fig:cumulative_skew_kurt} is in accordance with the observation that the market clustering relates to a relative increase of only the upward price fluctuations. However, this does not mean that the kurtosis results in \Cref{table:stability} are solely caused by the upper tail. The robustness checks in \Cref{table:rob_logreturn} for partial data show that the relation between market clustering and the kurtosis is also significant when the tail observations of the log return distributions are left out of the analyses. Furthermore, the lack of clear unconditional relation of price instability and market clustering in the negative tail does not preclude a possibility of a conditional relation. We investigate market conditions as a possible confounding factor in the panel data framework.

The significance of the relation between market clustering and price instability varies over time, as the test results for shorter time spans indicate. \Cref{table:2month} repeats the results of Panel a) in \Cref{table:stability} for a time window of two months. Approximately half of the kurtosis test results for a time window of two months are the same as in the yearly results. For 2009, \Cref{table:2month} shows a clear positive relation between the kurtosis and market clustering. During the period 2010-2011 the positive relation seems to apply to the end of 2010 and the first half of 2011. In 2012 and the first half of 2013 no consistent relation exists for any of the measures or time window. For the end of 2013 until the end of the sample the kurtosis results are mostly positive. The significance of the results at shorter time scales is reduced because the time series measures have a higher spread at shorter time scales, while the number of observations stays the same. The significance of the relation between market clustering and price instability might vary because the samples within the time windows are too small. Nevertheless, the relation between market clustering and the kurtosis is positive in more than half the test statistics for the two month time windows.

\begin{table}[!htb]
    \centering
    \resizebox*{\textwidth}{!}{
\begin{tabular}{rcccccccccccc}
\toprule
 & \multicolumn{6}{c}{2009} & \multicolumn{6}{c}{2010} \\
 \cmidrule(lr){2-7} \cmidrule(lr){8-13}
 & 1 & 3 & 5 & 7 & 9 & 11 & 1 & 3 & 5 & 7 & 9 & 11 \\
\midrule
MAD & $==$ & $+=$ & $==$ & $==$ & $==$ & $=-$ & $--$ & $=-$ & $++$ & $++$ & $++$ & $=+$ \\
Variance & $==$ & $++$ & $++$ & $=+$ & $++$ & $==$ & $==$ & $==$ & $++$ & $++$ & $++$ & $++$ \\
Skewness & $==$ & $==$ & $==$ & $++$ & $==$ & $==$ & $+=$ & $==$ & $==$ & $==$ & $++$ & $++$ \\
Kurtosis & $++$ & $==$ & $++$ & $++$ & $++$ & $+=$ & $+=$ & $++$ & $==$ & $==$ & $++$ & $++$ \\
 & \multicolumn{6}{c}{2011} & \multicolumn{6}{c}{2012} \\
 \cmidrule(lr){2-7} \cmidrule(lr){8-13}
 & 1 & 3 & 5 & 7 & 9 & 11 & 1 & 3 & 5 & 7 & 9 & 11 \\
\midrule
MAD & $==$ & $==$ & $==$ & $++$ & $==$ & $++$ & $==$ & $==$ & $==$ & $==$ & $==$ & $==$ \\
Variance & $==$ & $==$ & $==$ & $++$ & $++$ & $++$ & $==$ & $+=$ & $==$ & $==$ & $==$ & $+=$ \\
Skewness & $++$ & $==$ & $++$ & $++$ & $++$ & $==$ & $==$ & $==$ & $==$ & $==$ & $==$ & $==$ \\
Kurtosis & $++$ & $++$ & $++$ & $+=$ & $==$ & $==$ & $==$ & $==$ & $==$ & $==$ & $==$ & $==$ \\
 & \multicolumn{6}{c}{2013} & \multicolumn{6}{c}{2014} \\
 \cmidrule(lr){2-7} \cmidrule(lr){8-13}
 & 1 & 3 & 5 & 7 & 9 & 11 & 1 & 3 & 5 & 7 & 9 & 11 \\
\midrule
MAD & $++$ & $==$ & $==$ & $=+$ & $==$ & $++$ & $++$ & $++$ & $++$ & $++$ & $++$ & $==$ \\
Variance & $++$ & $++$ & $++$ & $++$ & $+=$ & $++$ & $++$ & $++$ & $++$ & $++$ & $++$ & $+=$ \\
Skewness & $==$ & $==$ & $+=$ & $==$ & $++$ & $==$ & $++$ & $++$ & $==$ & $++$ & $+=$ & $+=$ \\
Kurtosis & $==$ & $+=$ & $++$ & $==$ & $++$ & $++$ & $++$ & $++$ & $=+$ & $++$ & $++$ & $==$ \\
\bottomrule
\end{tabular}
    }
    \caption{Testing for a relation between market clustering and price instability -- 2 month window. \\
    { \footnotesize Repetition of Panel a) in \Cref{table:stability} for time windows of two months instead of one year. Contrary to \Cref{table:stability}, here the critical value is 0.05. The dates in the first line indicate the first month of each time window.} }
    \label{table:2month}
\end{table}

The results for the skewness, kurtosis and outlier count are normalized by the volatility. We show the relation between the variance and market clustering separately in \Cref{table:stability,table:2month} and \Cref{fig:cumulative_skew_kurt}. In addition, we analyze the results for the MAD. We find no consistent relation between market clustering and the yearly MAD. We find a weak but consistent positive relation between market clustering and the yearly variance. \Cref{fig:cumulative_skew_kurt} shows that the discrepancy between the distributions is smaller for the MAD and variance than for the kurtosis. The results for time spans of two months (see \Cref{table:2month}) show an increase of the MAD and variance during the periods where the kurtosis results are consistently positive. The relation between market clustering and the MAD and variance is not informative in itself, since the stocks are traded in different markets. The observation that more (less) market clustering relates to stronger (weaker) price fluctuations is in accordance with the observation that market clustering relates relatively more to the variance than the MAD, because the MAD is more robust to outliers than the variance. Market clustering relates also to price instability measured relative to time-varying volatility. \Cref{table:rob_garch} shows the relation between market clustering and the yearly kurtosis of log returns normalized by the conditional standard deviation estimated by various GARCH models. This indicates that the relation between market clustering and price instability is not confined to periods of high volatility.

Using the partial coverage of our data set we can dispel concerns over reversed causation. Rather than market clustering causing price instability, unstable stocks might attract traders that prefer to trade in clusters. If the latter holds, then the relation between kurtosis and market clustering would be independent of what percentage of the total turnover traded is included in the data set. \Cref{table:turn-0.1} shows that the relation between market clustering and the kurtosis vanishes for stocks that are mainly traded by investors which are not included in the MiFID data set. The relation between the kurtosis and price instability is (not) significant for stocks with a high (low) percentage of the turnover traded within the data set. By difference-in-differences logic these results indicate that market clustering leads to price instability and not the other way round.

\begin{table}[!htb]
\centering
\begin{tabular}{rcccccc}
\hline
 & 2009 & 2010 & 2011 & 2012 & 2013 & 2014 \\
\hline
Mean & $==$ & $=-$ & $=-$ & $==$ & $-=$ & $==$ \\
Variance & $==$ & $==$ & $==$ & $==$ & $==$ & $=+$ \\
Skewness & $==$ & $==$ & $==$ & $==$ & $=-$ & $==$ \\
Kurtosis & $+=$ & $==$ & $==$ & $==$ & $==$ & $=-$ \\
\hline
\end{tabular}

\caption{Relation between market clustering and price instability for stocks mainly traded by investors outside MiFID data set. \\ {\footnotesize Repetition of Panel a) in \Cref{table:stability} for the stocks for which less than 10\% of the total yearly turnover is covered by the investors in the MiFID data set. This category contains on average 434 stocks, which is 44.5\% of the total group of selected stocks.}}
\label{table:turn-0.1}
\end{table}

\FloatBarrier

\subsection{Drivers of market clustering}\label{subsection:driversres}

An important question is whether our proposed clustering measure actually captures new, previously ignored information. We set up a dynamic panel data framework to investigate observable determinants of market clustering. Selected models, shown in \Cref{table:clust1all100}, suggest that clustering is a quite persistent process (if at time $t$ clustering is high (low), it is likely to be high (low) at $t+1$, too), mainly driven by commonalities, illiquidity, and size. Other stock specific variables have little to no effect in our setting. No more than $20\%$ of the clustering measure variation can be explained by considered characteristics that would proxy for investor preferences. Thus a large part of the clustering measure variation remains unexplained and is likely due to accidental portfolio overlap.

We estimate a variety of specifications, but choose to report only valid models (according to the Hansen-Sargan statistic and residual autoregression tests). We try to mitigate the instrument proliferation problem by collapsing GMM-style instruments as suggested by \cite{roodman_note_2009}. Using fewer lags for the instruments than all available leads the model to fail the Hansen-Sargan test of validity.

\begin{table}[!htb]
\centering
\resizebox*{!}{\dimexpr\textheight-6\baselineskip\relax}{
\begin{tabular}{llllll}
  \toprule
 & {Model 1} & {Model 2} & {Model 3} & {Model 4} & {Model 5}  \\ 
\midrule															
IV lags \\															
{ }{ }CLUST	&	$3:99$	&	$3:99$	&	$3:99$	&	$3:99$	&	$3:99$	\\
{ }{ }VaR\_	&	$1:99$	&	$1:99$	&	$1:99$	&	$1:99$	&	$1:99$	\\
{ }{ }VIX	&		          &	$3:99$	&	$3:99$	&		           &		\\
\midrule															
\multirow{2}{*}{$\text{CLUST}_{t-1}$}	&	0.103***	&	0.111***	&	0.114***	&	0.102***	&	0.101***	\\
	&	(0.014)	&	(0.015)	&	(0.008)	&	(0.014)	&	(0.014)	\\
\multirow{2}{*}{$\text{CLUST}_{t-2}$}	&	0.062***	&	0.074***	&	0.077***	&	0.061***	&	0.063***	\\
	&	(0.016)	&	(0.015)	&	(0.006)	&	(0.016)	&	(0.016)	\\
\multirow{2}{*}{VaR\_dev}	&	0.001	&	-0.056**	&	-0.023	&	0.014	&	-0.003	\\
	&	(0.026)	&	(0.023)	&	(0.019)	&	(0.026)	&	(0.025)	\\
\multirow{2}{*}{VaR\_chg}	&		&	0.018	&		&		&		\\
	&		&	(0.031)	&		&		&		\\
\multirow{2}{*}{MKTF}	&	0.053	&	0.113**	&	0.059	&	0.043	&	0.192***	\\
	&	(0.050)	&	(0.054)	&	(0.043)	&	(0.050)	&	(0.071)	\\
\multirow{2}{*}{MKTF$_{t-1}$}	&	0.175***	&	0.174***	&	0.128***	&	0.149***	&	0.174***	\\
	&	(0.045)	&	(0.045)	&	(0.039)	&	(0.050)	&	(0.046)	\\
\multirow{2}{*}{VIX}	&	0.178**	&	0.095	&	-0.038	&	0.122	&	0.237***	\\
	&	(0.073)	&	(0.070)	&	(0.076)	&	(0.093)	&	(0.076)	\\
\multirow{2}{*}{VIX$_{t-1}$}	&	0.380***	&	0.326***	&	0.359***	&	0.374***	&	0.343***	\\
	&	(0.099)	&	(0.103)	&	(0.079)	&	(0.098)	&	(0.100)	\\
\multirow{2}{*}{VIX$_{t-2}$}	&	-0.258***	&	-0.305***	&	-0.270***	&	-0.249***	&	-0.261***	\\
	&	(0.062)	&	(0.065)	&	(0.056)	&	(0.064)	&	(0.063)	\\

\midrule
\multicolumn{6}{l}{\textit{Results for MOM, MCAP and ILLIQ are on the next page.}}\\															
															
\midrule
No. IVs	&	172	&	316	&	246	&	178	&	178		\\							Sargan stat	&	177.475	&	242.396	&	239.030	&	175.007	&	185.105	\\
{ }{ }DF	&	158	&	301	&	229	&	161	&	161	\\
{ }{ }p-value	&	0.138	&	0.994	&	0.311	&	0.203	&	0.094	\\
AR($1$)	&	0.000	&	0.000	&	0.000	&	0.000	&	0.000	\\
AR($2$)	&	0.719	&	0.528	&	0.459	&	0.716	&	0.665	\\
corr$^2(y,\hat{y})$	&	0.181	&	0.188	&	0.193	&	0.174	&	0.181	\\

\bottomrule															
\end{tabular}															

\caption{Estimation results of dynamic panel data models for the clustering measure. \\
{\footnotesize This table contains estimation results of \Cref{eq:dynamicpanel} using a two-steps system GMM approach with collapsed GMM-style instruments. The dependent variable is the clustering measure. corr$^2_{y,\hat{y}}$ measures squared correlation between the dependent variable and the fitted values by the model. Coefficients for PB3, DY, and LEV3 are insignificant in all models and are not reported to conserve space. Here CLUST is multiplied by $100$ and $N=27031$. Standard errors are in parentheses below the estimates. Coefficients significant at $10$, $5$, and $1\%$ level are marked with *, **, and ***, respectively. Obvious subscripts $i$ and $t$ are omitted for brevity.} }
\label{table:clust1all100}
}
\end{table}

\FloatBarrier

\begin{table}[!htb]
\centering
\resizebox{\columnwidth}{!}{
\begin{tabular}{lllllll}
\toprule													
	&	Model 1	&	Model 2	&	Model 3	&	Model 4	&		&	Model 5	\\
\midrule													
\multirow{2}{*}{MOM}	&	-0.006	&	-0.11	&		&		&		&		\\
	&	(0.111)	&	(0.115)	&		&		&		&		\\
\multirow{2}{*}{MCAP}	&	-0.146	&	0.320*	&		&		&		&		\\
	&	(0.187)	&	(0.192)	&		&		&		&		\\
\multirow{2}{*}{ILLIQ}	&	0.186*	&	0.408***	&		&		&		&		\\
	&	(0.096)	&	(0.098)	&		&		&		&		\\
\multirow{2}{*}{MOMhigh}	&		&		&	-0.180	&	-0.081	&	\multirow{2}{*}{MOMup}	&	0.046	\\
	&		&		&	(0.151)	&	(0.153)	&		&	(0.122)	\\
\multirow{2}{*}{MOMlow}	&		&		&	0.037	&	0.05	&	\multirow{2}{*}{MOMdown}	&	-0.056	\\
	&		&		&	(0.095)	&	(0.127)	&		&	(0.134)	\\
\multirow{2}{*}{MCAPhigh}	&		&		&	0.750***	&	-0.011	&	\multirow{2}{*}{MCAPup}	&	-0.439**	\\
	&		&		&	(0.237)	&	(0.299)	&		&	(0.180)	\\
\multirow{2}{*}{MCAPlow}	&		&		&	0.416**	&	0.003	&	\multirow{2}{*}{MCAPdown}	&	-0.131	\\
	&		&		&	(0.174)	&	(0.199)	&		&	(0.154)	\\
\multirow{2}{*}{ILLIQhigh}	&		&		&	0.198	&	-0.079	&	\multirow{2}{*}{ILLIQup}	&	0.111	\\
	&		&		&	(0.127)	&	(0.143)	&		&	(0.116)	\\
\multirow{2}{*}{ILLIQlow}	&		&		&	0.499***	&	0.303***	&	\multirow{2}{*}{ILLIQdown}	&	0.252**	\\
	&		&		&	(0.089)	&	(0.107)	&		&	(0.100)	\\
\bottomrule													
\end{tabular}													

\caption*{\Cref{table:clust1all100}: \textit{continued}. \\ {\footnotesize Variables with suffixes 'high', 'low', 'up', and 'down' are interacted with $\mathbbm{1}_{(\text{VIX} \geq 25)}$, $\mathbbm{1}_{(\text{VIX}<25)}$,  $\mathbbm{1}_{(\text{MKTF} \geq 0)}$, and $\mathbbm{1}_{(\text{MKTF}<0)}$, respectively.}}
}
\end{table}

\Cref{table:clust1all100} demonstrates that crowded trading is a persistent feature as the clustering measure exhibits significant positive dependence on the lagged values of market clustering in all models. Herding, lasting for at least multiple months in upward markets, could be one of the mechanisms related to clustering. In the cases where market clustering results from accidental portfolio overlaps, continuing clustering may be observed due to spreading the orders over time to reduce market impact. The persistence of market clustering suggests the need for further research with adjusted measures of market clustering, that differentiates between buy and sell orders. Furthermore, investigation of the stability of the investors’ pools involved in clustered trades would be helpful in understanding the effects of market clustering.

There is little evidence that individual downside risk affects the clustering measure. From all of the five models in \Cref{table:clust1all100}, the only significant coefficient is for the deviation from the median VaR (VaR\_dev) in Model $2$, but there is no solid reason to conclude that this model is superior to others. On the contrary, this model employs the largest set of instruments and could be the most vulnerable to biases. All of these models consider clustering measure as endogeneous variable in line with our hypothesis that market clustering causes price instability.

Market direction and market risk affect market clustering in multiple ways. First, all models indicate that increase (decrease) in market returns or market volatility in the previous month lead to significantly more (less) clustering per average stock. Second, lagged general market uncertainty (VIX$_{t-2}$) has a negative effect. We interpret this as a short-term corrective mechanism: when increased market volatility leads to more crowded trades, then a month afterwards the trading subsides (because the funds are used up, the interest is transferred elsewhere, investors get scared of continuing uncertainty, or some other reason) and so do the clustered activities. Third, market conditions play a role through asymmetric effects of stock size and illiquidity on clustering measure. Models $3$ and $4$ look at the effect for high and low volatility states, and Model $5$ shows differences across up and down markets. The specifics of these asymmetries and implications are discussed in the last paragraph of this section.

Models $2$ and $3$ also add VIX to the GMM-style instruments thus allowing to correct for potential VIX endogeneity, i.e., that clustering feeds aggregate market volatility. By comparing Models $1$ and $2$, we note that lags of VIX among instruments matter a lot for some coefficients. The contemporaneous effect of MKTF increases and becomes significant, while contemporaneous effect of VIX becomes insignificant. The coefficients for lagged commonalities hardly change. Further, in Model $3$ MKTF$_t$ is again insignificant even though lagged VIX values serve as instruments. This may be due to the fact that Model $2$ incorporates VaR\_chg, also among instruments, which is correlated with MKTF ($-0.203$ as seen in \Cref{table:corrs}). In Model $4$ only lags of two variables are considered as instruments, but VIX$_t$ has an insignificant coefficient likely because asymmetric effects with respect to the level of volatility are incorporated. Similarly MFTK$_t$ becomes significant in Model $5$ when asymmetric effects with respect to the market direction play a role. The exact effect of current MKTF and VIX on crowding is unclear.

Illiquidity and size are the only stock-specific variables that affect clustering, while momentum, price-to-book ratio, dividend yield and leverage do not yield a significant coefficient in any of the models. To better understand illiquidity and size effects, we investigate asymmetries across market conditions (Models $3-5$). We find that in quiet times market participants tend to cluster around less liquid stocks (significant coefficients for ILLIQlow and insignificant for ILLIQhigh), perhaps because they are willing to take more risks. Less liquid stocks attract more clustered trades in downward markets, too (Model $5$). This result is consistent with fire sales. Market capitalization (MCAP) has (marginally) significant coefficients in Models $2$ and $3$, when VIX is treated as endogeneous. In high volatility markets large firms attract more crowded attention than they do in low volatility markets (coefficients of $0.750$ vs. $0.416$). Large stocks are frequently dividend paying, are likely index constituents, and are considered less risky, thus such trading behavior may be viewed as a flight-to-safety within equities. In upward markets MCAP has a negative coefficient (Model $5$) -- small firms get the attention which is consistent with efforts to reap size premium.

To summarize the insights from this section, the results support theories that market clustering could be a consequence of multiple mechanisms. For one, herding induces persistence in the clustering measure time series. Next, willingness to take up more risks in low volatility markets and fire sales in high volatility markets both manifest as increased clustering around less liquid stocks. Finally, depending on market conditions, we see market clustering resulting from two size related phenomena -- flight-to-safety and reaping size premium. Interestingly, we see no evidence that stock selection based on considered fundamental characteristics would lead to market clustering.

\subsection{Downside risk, upside potential, and clustering}\label{subsection:varres}

We now turn to the causal analysis of market clustering and price instability. We employ a dynamic panel data model to analyze whether our newly proposed measure actually has additional explanatory power in modeling changes in downside risk and upside potential in addition to all commonly used conditioning variables (as discussed in \Cref{section:data}). In short: we find that market clustering indeed causes price instability, but not all the time.

Finding valid models for changes in the left and right $5\%$ quantiles is not trivial. Model $1$ is valid, but even with strict lag selection for the GMM-style instruments the total number of instruments is roughly twice as large as the cross-section. Model $2$ uses all possible lags for dependent variable and clustering measure but Sargan's test rejects the validity of collapsed instruments. We do not discuss these two models any further as they are likely unreliable. Finally, in Models $3-6$, we allow VIX into the GMM-style instrument set and obtain statistically valid models. All of the models consider price instability measure as an endogenous variable in line with our concerns that price instability could lead to market clustering. Models $3$ and $4$ look at changes in downside risk, Models $5$ and $6$ look at changes in upside potential.

Consistent with the outcome of stochastic dominance analysis, there is no causal relation between CLUST and price instability in the negative tail (Model $3$). Model $4$, however, reveals that in high volatility markets the relation is significant. This makes crowded trading a dangerous phenomena, likely fostering contagion. On the positive side of the return distribution (Models $5$ and $6$), clustering leads to price instability in both high and low volatility periods. Based on the squared correlation between the dependent variable and fitted values, the positive tail is harder to explain, nonetheless. Lagged CLUST yields insignificant coefficients in all models thus the direct causation is contemporaneous and does not extend to the next period.

Other coefficients have the expected signs or are insignificant. Strongly significant variables come from two categories: aggregate market related (MKTF and VIX) and derived from returns (MOM, MCAP, and ILLIQ). Upward movement and trend in the market index leads to smaller individual risks (thus negative changes in VaR) and more gradual price increases (thus slightly negative changes in VLuck). Current increase in volatility also increases changes in downside risk. For the positive tail of the distribution we again see the short-term corrective mechanism: higher volatility at time $t$ implies higher average change in upside potential, but taking advantage of this will result in reducing the upside potential for the next period. Positive momentum, higher market capitalization, and higher illiquidity have consistently negative effect on the changes in log return distribution quantiles.  Fundamental characteristics (PB3, DY, and LEV3) do not contribute to explaining the time variation in positive and negative quantiles of return distribution.

All in all, we show that market clustering causes contemporaneous price instability. The relation is present in the negative tail only during turmoil and in the positive tail independent of volatility level.

\begin{table}[!htb]
\centering
\resizebox{\columnwidth}{!}{
\begin{tabular}{lllllll}
\toprule													
	&	Model 1	&	Model 2	&	Model 3	&	Model 4	&	Model 5	&	Model 6	\\
\midrule													
N	&	27617	&	27295	&	27295	&	27295	&	27295	&	27295	\\
GMM IV lags\\
{ }{ }dep.var.	&	3:5	&	2:99	&	2:99	&	2:99	&	2:99   &	2:99	\\
{ }{ }CLUST	&	1:3	&	2:99	&	2:99	&	2:99	&	2:99	&	2:99	\\
{ }{ }VIX	&		&		&	2:99	&	2:99	&	2:99	&	2:99	\\
Collapse?	&	FALSE	&	TRUE	&	TRUE	&	TRUE	&	TRUE	&	TRUE	\\
\midrule													
\multirow{2}{*}{y$_{t-1}$}	&	0.140***	&	0.149***	&	0.136***	&	0.131***	&	0.103***	&	0.100***	\\
	&	(0.013)	&	(0.015)	&	(0.012)	&	(0.013)	&	(0.017)	&	(0.017)	\\
\multirow{2}{*}{y$_{t-2}$}	&	0.095***	&	0.109***	&	0.095***	&	0.090***	&	0.073***	&	0.071***	\\
	&	(0.012)	&	(0.013)	&	(0.011)	&	(0.011)	&(0.011)		&(0.011)		\\
\multirow{2}{*}{CLUST} &	0.298	&	7.925***	&	1.598	&		&	6.150***	&		\\
	&	(0.531)	&	(2.004)	&	(1.209)	&		&	(1.462)	&		\\
\multirow{2}{*}{CLUST$_{t-1}$}	&		&	0.076	&	0.179	&		&	-0.192	&		\\
	&		&	(0.506)	&	(0.401)	&		&	(0.387)	&		\\
\multirow{2}{*}{CLUSTlow}	&		&		&		&	1.398	&		&	3.055**	\\
	&		&		&		&	(1.352)	&		&	(1.44)	\\
\multirow{2}{*}{CLUSTlow$_{t-1}$}	&		&		&		&	-0.079	&		&	-0.001	\\
	&		&		&		&	(0.469)	&		&	(0.481)	\\
\multirow{2}{*}{CLUSThigh}	&		&		&		&	5.315**	&		&	7.341***	\\
	&		&		&		&	(2.176)	&		&	(2.121)	\\
\multirow{2}{*}{CLUSThigh$_{t-1}$}	&		&		&		&	1.239	&		&	-0.214	\\
	&		&		&		&	(0.975)	&		&	(0.701)	\\

\bottomrule													

\end{tabular}

\caption{Estimation results of dynamic panel models for VaR\_chg and VLuck\_chg. \\
{\footnotesize This table contains estimation results of \Cref{eq:dynamicpanel} using a two-steps system GMM approach.  Models $1-4$ use $y_{it}= \operatorname{VaR\_chg}_{it}$ and Models $5-6$ use $y_{it}= \operatorname{VLuck\_chg}_{it}$. Models $4$ and $6$ introduce $\text{CLUSTlow}=\text{CLUST} \times \mathbbm{1}_{(\text{VIX}<25)}$ and $\text{CLUSThigh}=\text{CLUST} \times \mathbbm{1}_{(\text{VIX}\geq 25)}$ to account for asymmetric effects. corr$^2_{y,\hat{y}}$ measures squared correlation between the dependent variable and the fitted values by the model. Standard errors are in parentheses below the estimates. Coefficients significant at $10$, $5$, and $1\%$ level are marked with *, **, and ***, respectively. Obvious subscripts $i$ and $t$ are omitted for brevity.} }
\label{table:dynfull12tails}
}
\end{table}

\FloatBarrier

\begin{table}[!htb]
\centering
\resizebox{\columnwidth}{!}{
\begin{tabular}{lllllll}
\toprule													
	&	Model 1	&	Model 2	&	Model 3	&	Model 4	&	Model 5	&	Model 6	\\
\midrule
\multirow{2}{*}{MKTF}	&	-0.290***	&	-0.286***	&	-0.303***	&	-0.307***	&	-0.047***	&	-0.050***	\\
	&	(0.015)	&	(0.019)	&	(0.016)	&	(0.016)	&	(0.016)	&	(0.015)	\\
\multirow{2}{*}{MKTF$_{t-1}$}	&	-0.149***	&	-0.154***	&	-0.144***	&	-0.146***	&	-0.107***	&	-0.105***	\\
	&	(0.010)	&	(0.012)	&	(0.010)	&	(0.010)	&	(0.011)	&	(0.011)	\\
\multirow{2}{*}{VIX}	&	0.084***	&	0.065**	&	0.117***	&	0.095***	&	0.158***	&	0.141***	\\
	&	(0.023)	&	(0.028)	&	(0.024)	&	(0.027)	&	(0.024)	&	(0.026)	\\
\multirow{2}{*}{VIX$_{t-1}$}	&	-0.017	&	-0.031	&	0.009	&	0.01	&	-0.113***	&	-0.107***	\\
	&	(0.021)	&	(0.026)	&	(0.023)	&	(0.024)	&	(0.023)	&	(0.024)	\\
\multirow{2}{*}{MOM}	&	-0.325***	&	-0.314***	&	-0.303***	&	-0.301***	&	-0.416***	&	-0.412***	\\
	&	(0.030)	&	(0.039)	&	(0.030)	&	(0.030)	&	(0.036)	&	(0.034)	\\
\multirow{2}{*}{MCAP}	&	-0.168***	&	-0.136***	&	-0.337***	&	-0.294***	&	-0.213***	&	-0.175***	\\
	&	(0.024)	&	(0.035)	&	(0.038)	&	(0.043)	&	(0.046)	&	(0.049)	\\
\multirow{2}{*}{ILLIQ}	&	-0.011	&	-0.032*	&	-0.060***	&	-0.049***	&	-0.046***	&	-0.027	\\
	&	(0.010)	&	(0.016)	&	(0.015)	&	(0.016)	&	(0.018)	&	(0.017)	\\
\multirow{2}{*}{PB3}	&	0.030	&	0.036	&	0.008	&	0.013	&	0.087	&	0.085	\\
	&	(0.051)	&	(0.065)	&	(0.053)	&	(0.051)	&	(0.056)	&	(0.053)	\\
\multirow{2}{*}{DY}	&	0.027	&	0.019	&	0.016	&	0.019	&	0.013	&	0.022	\\
	&	(0.017)	&	(0.023)	&	(0.023)	&	(0.022)	&	(0.033)	&	(0.029)	\\
\multirow{2}{*}{LEV3}	&	-0.004	&	-0.002	&	-0.008**	&	-0.007**	&	-0.001	&	0.000	\\
	&	(0.003)	&	(0.006)	&	(0.003)	&	(0.003)	&	(0.004)	&	(0.003)	\\

\midrule
No. IVs	&	603	&	170	&	241	&	316	&	241	&	316	\\							Sargan stat	&	241.224	&	207.707	&	239.556	&	242.341	&	240.421	&	237.452	\\
{}{}DF	&	590	&	156	&	227	&	300	&	227	&	300	\\
{}{}p-value	&	1.000	&	0.004	&	0.271	&	0.994	&	0.258	&	0.997	\\
AR($1$)	&	0.000	&	0.000	&	0.000	&	0.000	&	0.000	&	0.000	\\
AR($2$)	&	0.440	&	0.265	&	0.265	&	0.146	&	0.355	&	0.504	\\
corr$^2_{y,\hat{y}}$	&	0.158	&	0.109	&	0.151	&	0.150	&	0.070	&	0.083	\\
\bottomrule													

\end{tabular}
\caption*{\Cref{table:dynfull12tails}: \textit{continued}.}
}
\end{table}

\section{Discussion}
	\label{section:discussion}

We have shown some suggestive evidence for a causal relation between market clustering and price instability on the individual stock level. There seems to be a consistent and robust positive relation between market clustering and the kurtosis, the skewness, the positive tail index, the positive outlier count, and the right 5\% quantile of the log return distribution. The positive relation between market clustering and the left 5\% quantile of the log return distribution is conditional on periods of high volatility. Focusing on extreme price fluctuations, that is the tails of the normalized log return distribution, we find that market clustering generally causes an increase of large upward price shocks. Increases of large downward shocks due to market clustering turns out to be present only in financial turmoil. Findings on the positive tail are consistent with herding, while findings on the negative tail are consistent with fire sales.

We also provide some insights into investor behavior that likely leads to market clustering. The persistence of our market clustering measure could be explained by herding and order spreading over time. Market conditions obviously affect trading decisions. We find an indication that the homogeneity of the investors' pool per stock increases if there is a positive trend in the market or increase in aggregate volatility. However, the volatility effect is short-term and reverses in the month afterwards. Furthermore, we find asymmetries across market conditions. In quiet times investors prefer less liquid stocks. Consistent with fire sales, less liquid stocks are also traded by more homogeneous groups in downward markets. We discover behavior that is consistent with flight-to-safety within equities in the sense that in high volatility markets large firms attract more crowded attention. In upward markets, on the contrary, investors are interested in size premium from small firms.

Our analysis contributes to the existing literature on three levels. Firstly, we study the influence of trading behavior on price dynamics using novel granular trading data. To our knowledge the MiFID data set has not been used for this type of market microstructure research before. Secondly, the idea and method to measure market clustering and its impact on price instability are new to market microstructure research. The use of complex network theory makes the method suitable for large-scale data. The methodological framework can be extended to study the effects of any feature of the market microstructure. Thirdly, the main contribution is the indication of a causal relation between the market clustering and price instability shown in a dynamic panel data model. 


The use of network theory in identifying meaningful motifs in market microstructure research is promising because the model is applicable to all types of market microstructure patterns. Firstly, the influence of trading behaviour on price dynamics can be investigated using other microstructure motifs, for example the influence of the diversification of the investors on the price dynamics of the traded stocks. Differentiation between buy and sell orders would enhance the understanding of the difference in dependence between the positive and negative tail of the price dynamics. The persistence of the market clustering measure -- evident in consistent positive dependence on past, lagged values of market clustering -- is worthy of further investigation of the time-dependence of the configuration of the investors’ pools involved in clustered trades. Secondly, the method can be used for portfolio holdings data and could, for example, contribute to the literature on price comovements due to common active mutual fund owners \citet{anton2014}. Thirdly, the method can be used to study trading patterns separate from price dynamics, for example, the evolution of clustering patterns over time.

Further investigation is needed to analyze what fuels clustering measure persistence. Also, the relation between clustering and current market conditions needs further attention, for example, what is the mechanism of spillovers in each case?


\clearpage



\appendix
\newpage
	
\section{Network reconstruction from the degree sequences}
	\label{appendix:networkreconstruction}

We define a bipartite network that describes the aggregate trading behaviour during a particular month.
For each month $t$ the investment behaviour comprised by the data is represented in a binary bipartite graph.
The bipartite network has two layers with edges between the two layers. No edges occur between two nodes in the same layer. The nodes of the first layer represent the set of firms $F_t$ that perform trades during month $t$.
The nodes of the second layer represent set of the securities $S_t$ that are traded during month $t$.
The individual firms are indicated by the number $f$ each month separately, such that $[\text{firm}~f] \in F_t$.
The indicator can take the values $f=1,...,n_{F,t}$ with $n_{F,t}$ the number of elements in $F_t$.
The individual securities are indicated by the number $s$ each month separately, such that $[\text{security}~s] \in S_t$.
The indicator can take the values $s=1,...,n_{S,t }$ with $n_{S,t}$ the number of elements in $S_t$.

The performed transactions are represented by the edges between the firms and the securities, denoted by the binary adjacency matrix $A_t$ with elements $a_{fs,t}$. The size of matrix $A_t$ is $n_{F,t}\times n_{S,t}$. The transactions are represented as follows: $a_{fs,t}=1$ in case firm $f$ traded in security $s$ during month $t$ and $a_{fs,t}=0$ otherwise.
The degree $d_{f,t}$ of firm $f$ during month $t$ is given by
\begin{equation}
d_{f,t}=\sum\limits_{s=1}^{n_{S_t}} a_{fs,t}
\end{equation}
and the degree of security $s$ is given by
\begin{equation}
d_{s,t}=\sum\limits_{f=1}^{n_{F_t}} a_{fs,t}.
\end{equation}
The set $D_{t,\text{obs}}$ contains the degrees of all nodes in month $t$, such that
\begin{equation}
d_{f,t},d_{s,t}\in D_{t,\text{obs}}\forall f,s
\end{equation}

The graph indicates only whether a trade of a firm in a security occurs. None of the following measures are represented: The number of transactions, the number of underlying securities or the turnover. Furthermore, the graph does not distinguish between buy and sell transactions or between agency and principal transactions.

The ensemble of graphs $\mathcal{G}$ contains all binary, bipartite networks where the number of elements in the top layer is $n_{F,t}$ and the number elements in the bottom layer is $n_{S,t}$. $\mathcal{G}$ is a grandcanonical ensemble, which means that all the graphs have the same number of nodes buth the number of links varies between zero and $n_{F,t}\times n_{S,t}$. The element $X\in\mathcal{G}$ is the configuration matrix of an arbitrary network configuration.
Given the number of firms $n_{F,t}$ and the number of securities $n_{S,t}$, there are $2^{n_{F,t}\times n_{S,t}}$ possible configurations.
Each configuration $X$ contains for each firm-security pair the information whether the firm trades the security ($a_{sf}(X)=1$) or does not trade the security ($a_{sf}(X)=0$). $X$ does not denote the observed configuration, but any arbitrary configuration such that $X\in\mathcal{G}$.

The goal is to find the probability distribution $P(X|D_{t,\text{obs}})$ over all possible trading configurations, which contains no other information than the observed degree sequences, i.e. the expected values of the degree sequences based on the probability distribution equal the observed degree sequences.
Besides the information of the degree sequences, all other information about the placing of the trades is removed, such that no other properties of the observed trading configuration can be reconstructed from the probability distribution.

We apply the maximum entropy method to the bipartite network analogously to applications to monopartite networks \citet{newman2004}. The 'degree of disorder' in the system defined by the probability distribution $P(X)$ over all network configurations $X\in\mathcal{G}$ is given by Shannon's entropy
\begin{equation}
S = - \sum\limits_{X\in\mathcal{G}}P(X)\ln P(X).
\end{equation}
Shannon's entropy can be seen as the weighted average of the amount of information. For example, in case $N$ events are equally probable $P(X)=1/N\rightarrow-\ln P(X)=\ln N$, where $\ln N$ is the minimum number of decisions leading to the information which event happens (in the natural unit of information).
In case of no constraints, Shannon's entropy would be maximized when each configuration $X$ occurs with equal probability.
We find the probability distribution $P(X|D_{t,\text{obs}})$ by maximizing Shannon's entropy numerically under the constraints of the degree sequences. By maximisation of the entropy under the constraints, all information except the constraints is randomized.
The aimed probability distribution $P(X|D_{t,\text{obs}})$ contains all information that can be possibly known based on the degrees of all nodes, but all other information is excluded.

In order to approximate the distribution $P(X|D_{t,\text{obs}})$ that excludes all information but the degrees of the nodes, Shannon's entropy is maximized under the constraint that the expected value of the degree sequences in $P(X|D_{t,\text{obs}})$ equal the observed degree sequences:
\begin{equation}
\langle d_f \rangle = d_{f,t,\text{obs}}, ~~\langle d_s \rangle = d_{s,t,\text{obs}} ~~\forall f,s.
\end{equation}
In total there are $n_{F}+n_{S}+1$ constraints:
\begin{equation}
\begin{cases}
d_{f,t,\text{obs}} = \sum\limits_{X\in\mathcal{G}}P(X)d_f(X)~~~\forall f,
\\
d_{s,t,\text{obs}} = \sum\limits_{X\in\mathcal{G}}P(X)d_s(X)~~~\forall s,
\\
1 = \sum\limits_{X\in\mathcal{G}}P(X),
\end{cases}
\end{equation}
where the last expression is the normalization of the probability distribution.
We define the Lagrange multiplier to find the optimum:
\begin{equation}
\resizebox{\textwidth}{!}{$
L=
S
+\alpha\Big(1-\sum\limits_{X\in\mathcal{G}}P(X)\Big)
+\sum\limits_{f=1}^{n_{F,t}}
\beta_f
\Big
(d_{f,t,\text{obs}} - \sum\limits_{X\in\mathcal{G}}P(X)d_f(X)\Big)
+\sum\limits_{s=1}^{n_{S,t}}
\beta_s\Big(d_{s,t,\text{obs}} - \sum\limits_{X\in\mathcal{G}}P(X)d_s(X)\Big).
$}
\end{equation}
The replacement
\begin{equation}
P'(X)= P(X)+\epsilon\eta(X),
\end{equation}
with $\eta$ an arbitrary function, allows to find the functional derivative
\begin{equation}
\sum\limits_{X\in\mathcal{G}}
\frac{\delta L}{\delta P(X)}
\eta(X)
=
\Big[
\frac{d}{d\epsilon}L(P'(X))
\Big]_{\epsilon=0}
,
\end{equation}
which results in
\begin{equation}
\frac{\delta L}{\delta P(X)}
=
\ln P(X) +
1+\alpha+
\sum\limits_{f=1}^{n_{F,t}}
\beta_f d_f(X)
+
\sum\limits_{s=1}^{n_{S,t}}
\beta_s d_s(X).
\end{equation}

Now the probability distribution $P(X|D_{t,\text{obs}}$ is determined by the optimum:
\begin{equation}
\frac{\delta L}{\delta P(X)}=0
~~\leftrightarrow~~
P(X|D_{t,\text{obs}})=\frac{e^{-H(X)}}{Z},
\end{equation}
with $H(X)$ the Hamiltonian
\begin{equation}
H(X)=\sum\limits_{f=1}^{n_{F,t}}
\beta_f d_f(X)
+
\sum\limits_{s=1}^{n_{S,t}}
\beta_s d_s(X),
\end{equation}
and $Z$ the partition function
\begin{equation}
Z=e^{1+\alpha}=\sum\limits_{X\in\mathcal{G}}e^{-H(X)}.
\end{equation}
The partition function can be written as
\begin{align}
Z&=\sum\limits_{X\in\mathcal{G}}
e^{-\sum\nolimits_f\beta_f \sum\nolimits_sa_{sf}(X)-\sum\nolimits_s\beta_s \sum\nolimits_fa_{sf}(X)}
\\&=\sum\limits_{X\in\mathcal{G}}
\prod\limits_{s,f}
e^{-\beta_f a_{sf}(X)-\beta_s a_{sf}(X)}
\\&=
\prod\limits_{s,f}\Big(1+
e^{-\beta_f -\beta_s }\Big).
\end{align}
Now we rewrite the equation, such that we make use of the mutual independence of probability distributions of all edges of the network:
\begin{align}
P(X|D_{t,\text{obs}})
&=
\prod\limits_{s,f}
P(a_{sf}(X)=1|D_{t,\text{obs}})^{a_{sf}(X)}P(a_{sf}(X)=0|D_{t,\text{obs}})^{1-a_{sf}(X)},
\end{align}
with
\begin{equation}
P(a_{sf}(X)=1|D_{t,\text{obs}}) = \frac{x_fx_s}{1+x_fx_s}
\end{equation}
and accordingly,
\begin{equation}
P(a_{sf}(X)=0|D_{t,\text{obs}}) = 1-P(a_{sf}(X)=1|D_{t,\text{obs}}) = \frac{1}{1+x_fx_s},
\end{equation}
following the notation (common in the literature \citet{newman2004}):
\begin{equation}
\begin{cases}
x_f=e^{-\beta_f}
\\
x_s=e^{-\beta_s}
\end{cases}.
\end{equation}
The variables $x_f$ and $x_s$ are called 'hidden variables'.
The result is a probability of a trading relation for each specific firm-security pair separately.

\newpage
\section{Stochastic dominance of price instability measures}

\begin{figure}[!htb]
\centering
    \includegraphics[width=.95\textwidth]{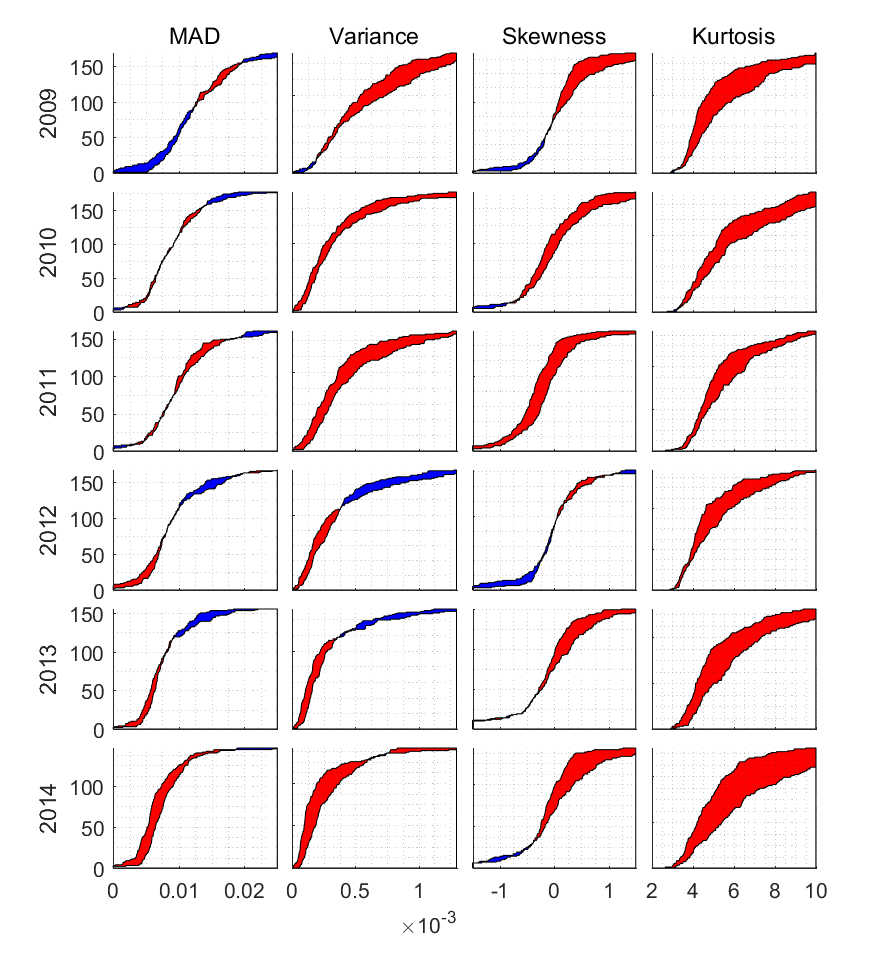}
    \caption{Cumulative distributions for low and high market clustering per time series measure (MAD, variance, skewness and kurtosis) and per year.\\
     {\footnotesize The space in between the distributions for low and high market clustering is colored to indicate which distribution is higher. Red means that distribution $H$ (high market clustering) exceeds distribution $L$ (low market clustering) and vice versa for blue.} }
    \label{fig:cumulative_skew_kurt}
\end{figure}

\begin{figure}[!htb]
    \centering
    \includegraphics[width=\textwidth]{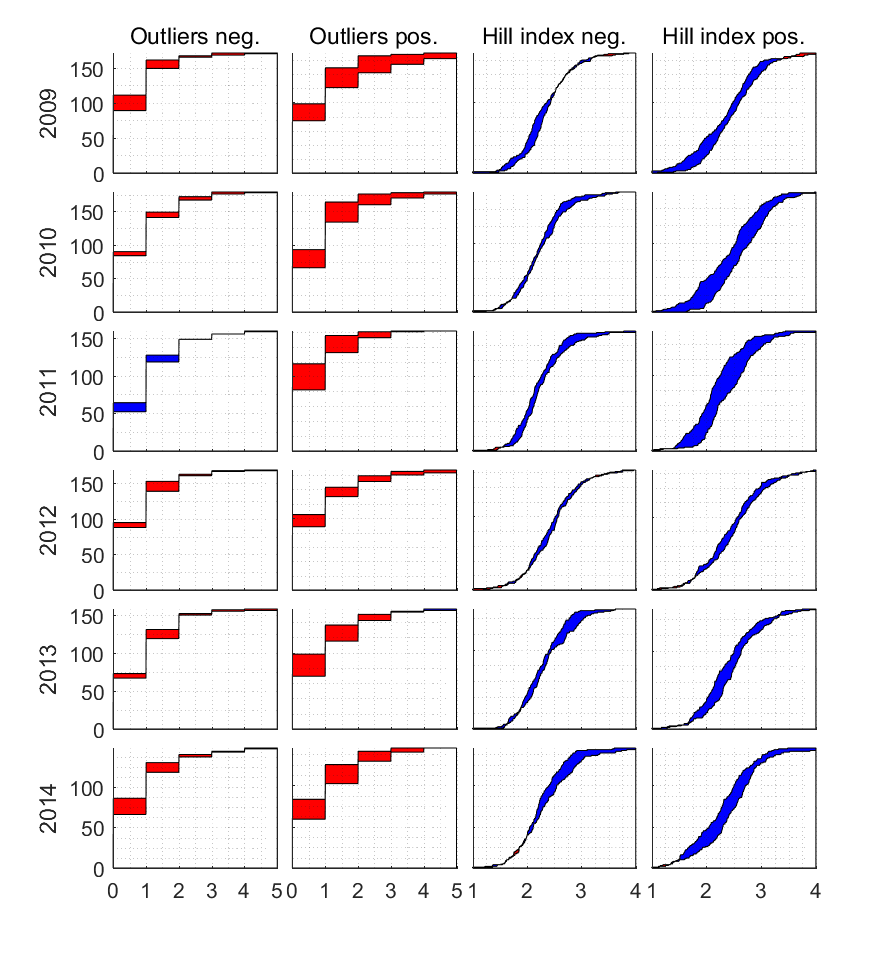}
    \caption{Cumulative distributions for low and high market clustering per time series measure (number of negative outliers, number of positive outliers, Hill index for the negative tail and hill index for the positive tail) and per year.\\
     {\footnotesize See \Cref{fig:cumulative_skew_kurt} for explanation.}}
    \label{fig:cumulative_hill_out}
\end{figure}

\section{Robustness checks}

\Cref{table:rob_logreturn} show the results for the relation between market clustering and the kurtosis for various segments of the log return distribution.
The results for the kurtosis do not depend in particular on the tails of the log return distribution.

\begin{table}[!htb]
\centering
\begin{tabular}{rcccccc}
\hline
 & 2009 & 2010 & 2011 & 2012 & 2013 & 2014 \\
\hline
$10-90$ & $++$ & $++$ & $++$ & $++$ & $++$ & $+=$ \\
$20-80$ & $++$ & $++$ & $++$ & $++$ & $++$ & $==$ \\
$30-70$ & $++$ & $++$ & $++$ & $++$ & $++$ & $==$ \\
$40-60$ & $++$ & $=+$ & $++$ & $=+$ & $++$ & $++$ \\
\hline
\end{tabular}

\caption{The kurtosis results from \Cref{table:stability} for partial data. \\
{\footnotesize Firstly, we order the log returns time series per stock and per year in ascending order. Secondly, we select the segments of the log return distribution as shown in the left column (in percentages). For example, the last line shows the results for the segment 40\%-60\% (the middle part), which means that we remove the first 40\% and the last 40\% of the orderd log return distribution.}}
\label{table:rob_logreturn}
\end{table}

\Cref{table:rob_marketclustering} show the relation between market clustering and the kurtosis for different cross-sections of the market clustering distribution.
The critical value is 0.025 in all tables. The results show no consistent variation over the different cross-sections.

\begin{table}[!htb]
\centering
\begin{tabular}{rcccccc}
\hline
 & 2009 & 2010 & 2011 & 2012 & 2013 & 2014 \\
\hline
0-10 and 90-100 & $++$ & $==$ & $+=$ & $++$ & $++$ & $++$ \\
10-30 and 70-90 & $++$ & $==$ & $++$ & $==$ & $=+$ & $++$ \\
0-50 and 50-100 & $++$ & $++$ & $+=$ & $++$ & $++$ & $++$ \\
20-50 and 50-80 & $==$ & $=+$ & $==$ & $==$ & $++$ & $=+$ \\
\hline
\end{tabular}

\caption{The relation between market clustering and the kurtosis for different cross-sections of the market clustering distribution. \\
{\footnotesize For comparison, \Cref{table:stability} shows the results for the highest 33\% and the lowest 33\% of the stocks, ranked according to their market clustering measure, i.e. the selected regions are 0-33\% and 67-100\%.}}
\label{table:rob_marketclustering}
\end{table}

\Cref{table:rob_garch} shows the result for normalization by the time-varying standard deviation, estimated by various GARCH-type models. Normalization by the time-varying volatility means that the weight of the price fluctuations in periods of high volatility is effectively reduced in favour of the weight of the price fluctuations during tranquil periods.
We estimate for each stock the conditional volatility time series for the complete log return time series at once instead of each year separately. The EGARCH model allows the sign and the magnitude of the log returns to have separate effects on the volatility. In the GJR-GARCH model the effects of the positive and negative log returns are estimated separately. The EGARCH models are exponential and therefore less sensitive to outliers than the GJR-GARCH models. The addition of extra lags allows the volatility to vary on both shorter and longer time scales. For all GARCH models we we assume conditional normal distribution for the error term: $\varepsilon_{s,t}\sim \mathcal{N}(0,\sigma^2_{s,t})$. This assumption is probably violated for some of the stocks. We assume that the consequences of this violation are limited.

The relation between market clustering and price instability remains consistently positive when we account for the time-varying volatility. We find no apparent variation of the results for the relation between market clustering and price instability for the different GARCH models for the conditional volatility. Apparently, market clustering causes downward price shocks not only during volatile periods but also when the price is more stable.

\begin{table}[!htb]
\centering
\begin{tabular}{rcccccc}
\hline
 & 2009 & 2010 & 2011 & 2012 & 2013 & 2014 \\
\hline
GARCH(1,1) & $++$ & $++$ & $==$ & $++$ & $++$ & $++$ \\
GARCH(2,2) & $++$ & $==$ & $==$ & $++$ & $++$ & $++$ \\
EGARCH(1,1) & $++$ & $++$ & $==$ & $++$ & $++$ & $++$ \\
GJR-GARCH(1,1) & $++$ & $++$ & $==$ & $++$ & $++$ & $++$ \\
\hline
\end{tabular}

\caption{The results for the yearly kurtosis of the log returns, normalized by the conditional standard deviation estimated by various GARCH models.}
\label{table:rob_garch}
\end{table}

\end{document}